\documentclass{aastex}  
\usepackage{emulateapj5,apjfonts}

\journalinfo{The Astrophysical Journal}

\slugcomment{Received 2001 June 26; Accepted 2001 August 2} 
 
\newcommand\ASCA{{\it ASCA}} 
\newcommand\BeppoSAX{{\it BeppoSAX}} 
 
\newcommand\HST{{\it HST}} 
\newcommand\ROSAT{{\it ROSAT}} 
\newcommand\RXTE{{\it RXTE}} 
\newcommand\FUSE{{\it FUSE}}
\newcommand\EUVE{{\it EUVE}}
\newcommand\Chandra{{\it Chandra}} 
\newcommand\kms{\ifmmode {\rm km\ s}^{-1} \else km s$^{-1}$\fi} 
\newcommand\ctssec{\ifmmode {\rm cts\ s}^{-1} \else cts s$^{-1}$\fi}  
\newcommand\ergsec{\ifmmode {\rm ergs\ s}^{-1} \else  
	ergs s$^{-1}$\fi}  
\newcommand\eflux{\ifmmode {\rm ergs\ s}^{-1}\;{\rm cm}^{-2} \else  
	ergs s$^{-1}$ cm$^{-2}$\fi}  
\newcommand\phflux{\ifmmode {\rm photons\ s}^{-1}\;{\rm cm}^{-2} 
	\else  	photons s$^{-1}$ cm$^{-2}$\fi}  
\newcommand\cc{\ifmmode {\rm cm}^{-3} \else cm$^{-3}$\fi}  
\newcommand\FWHM{\ifmmode {\rm FWHM} \else ${\rm FWHM}$\fi}  
\newcommand\Msun{\ifmmode M_{\odot} \else $M_{\odot}$\fi} 
\newcommand\Lsun{\ifmmode L_{\odot} \else $L_{\odot}$\fi} 
 
\newcommand\gtsim{\raisebox{-.5ex}{$\;\stackrel{>}{\sim}\;$}} 
\newcommand\Hbeta{\ifmmode {\rm H}\beta \else H$\beta$\fi} 
\newcommand\Kalpha{\ifmmode {\rm K}\alpha \else K$\alpha$\fi} 
\newcommand\NH{\ifmmode N_{\rm H} \else N$_{\rm H}$\fi} 
 
\shorttitle{Ton S180 \ASCA{} Observations} 
\shortauthors{Romano et al.\ 2001} 
 
\begin{document} 
	\title{A 12--day \mbox{\boldmath $ASC\!A$} Observation 
	of the Narrow-Line Seyfert 1 Galaxy Ton S180: 
	Time-Selected Spectroscopy} 
	\author{P.\ Romano\altaffilmark{1}, 
		T.J.\ Turner\altaffilmark{2,3}, S.\ Mathur\altaffilmark{1}, 
	  \&  	I.M.\ George\altaffilmark{2,3} }
\altaffiltext{1}{Department of Astronomy, Ohio State University,  
	140 West 18th Avenue, Columbus, OH  43210; 
	promano@astronomy.ohio-state.edu, mathur@astronomy.ohio-state.edu.}
\altaffiltext{2}{Joint Center for Astrophysics, 
	University of Maryland, Baltimore County,  
	Department of Physics, 1000 Hilltop Circle, Baltimore, 
	MD 21250; turner@lucretia.gsfc.nasa.gov, ian.george@gsfc.nasa.gov.} 
\altaffiltext{3}{Laboratory for High Energy Astrophysics, Code 660,
	NASA/Goddard Space Flight Center,
  	Greenbelt, MD 20771}

	\begin{abstract} 

We present an analysis of the X-ray variability properties of the
narrow-line Seyfert 1 galaxy \objectname[PHL 912]{Ton~S180}, 
based upon a 12-day continuous observation with \ASCA. 
Examination of the light curves reveals flux variations of  
a factor of 3.5 in the 0.7--1.3\,keV band and 3.9 in the 2--10\,keV band. 
Time-resolved spectroscopy, using approximately daily sampling,
reveals that the broad ``soft hump'' component at energies $<2$\,keV 
shows flux variations on timescales as short as 1 day that are well 
correlated with the photon index and the 2--10\,keV band flux. 
A broad Fe K$\alpha$ emission is detected. 
There is also a statistically significant evidence for a 
narrow Fe K$\alpha$ line at $\sim$ 6.8\,keV, indicating an origin 
in ionized material.  
We do not detect significant variations of the Fe K$\alpha$ line flux or 
equivalent width on timescales of $\sim$ 1 day--1 week. 

Despite evidence for correlated events in the power-law and 
soft hump on timescales of a day, the flux correlations clearly do 
not exist on all timescales. In particular, the softness ratio 
reveals spectral variability on timescales as short as
 $\sim 1000$\,s, indicating that the power-law continuum and soft 
hump fluxes are not well correlated on this timescale. 
The softness ratio also shows a slow decline across the observation, 
due to a combination of the different time-variability of the 
power-law continuum and soft hump flux on timescales of $\sim$ 1 week. 
 
Our timing analysis and time-selected spectroscopy indicate that 
the X-ray emission originates within 12 Schwarzschild radii.  
The amplitudes and timescales of the rapid variations we observed are 
consistent with those expected within disk-corona models.
Furthermore, the observed fast variability of the soft hump rules out 
an origin of the soft emission in large scale components, such as 
circumnuclear starburst. The $\Gamma$--soft hump correlation is 
consistent with the soft hump being produced by up-scattering 
of the accretion disk radiation within a patchy, flaring disk corona.

	\end{abstract} 
 
 
\section{Introduction}	 

The population of Seyfert 1 galaxies has a widely-used 
sub-classification into narrow-line Seyfert 1 galaxies (NLS1s) 
and broad-line Seyfert 1 galaxies (BLS1s). While the scheme appears 
to make an arbitrary distinction based primarily upon 
the widths of the optical emission lines (NLS1s having 
H$\beta \lesssim 2000$ ${\rm km\ s^{-1}}$, \citealt{Goodrich89}), 
this turns out to be an extremely useful terminology 
as the X-ray properties of the two sub-classes are systematically different.
Rapid and large-amplitude variability \citep[][hereafter BBF96; 
\citealt{Turnerea99b}]{BBF96} 
is a characteristic of NLS1s, with the excess variance 
\citep{Nandraea97a} typically an order of magnitude larger than that 
observed for samples of BLS1s  with the same luminosity distribution 
\citep{Turnerea99b,Leighly99I}. 
Spectral properties also vary across the Seyfert population with 
NLS1s showing  systematically steeper photon indices 
than those of BLS1s across both soft and hard X-ray bands, 
\citep{BBF96,BME97,TGN98,Leighly99II,Vaughan99b}. 
The anti-correlations between \Hbeta{} \FWHM{} 
and both X-ray spectral slope \citep{BBF96,Laorea97} and excess variance 
\citep{Turnerea99b} reveal a continuous range 
of parameter values between the NLS1 and BLS1 extremes of the population;    
the observed differences are now thought to be driven by a range in 
a fundamental physical parameter, such as the accretion rate. 

A popular idea, originally proposed by \citet{PDO95}, draws an 
analogy between Seyfert 1 galaxies and Galactic black-hole candidates. 
The latter show steepening in the high state. 
In this scenario, NLS1s are sources in the high-state, emitting at higher 
fractions of their Eddington luminosity, hence having higher fractional
accretion rates ($\dot{m} = \dot{M}/\dot{M}_{\mbox{\scriptsize Edd}}$). 
Given that NLS1s have comparable 
luminosity to that of the BLS1s, it has often been suggested that 
they also have 
relatively small central  black holes, i.e.,  $M_{\rm BH}
\approx 10^{6} \Msun$, as opposed to $M_{\rm BH} \approx 10^{8}  \Msun$,
which is typical for BLS1s  \citep[e.g.,][]{P2000,K2000}. 
Smaller black-hole masses naturally explain both the narrowness of
the optical emission lines, which are generated in clouds that have
smaller Keplerian velocities---hence smaller widths---and the extreme
X-ray variability, since the primary emission would originate in a
smaller region around the central engine \citep[e.g.,][]{Laorea97}.

Higher accretion rates lead to some observational predictions, 
such as a hotter accretion disk leading 
(via inverse Compton scattering) to enhanced soft X-ray emission 
and an ionized surface for the accretion disk \citep{MFR93}. 
Support for the ionized disk is found in the form of 
K$\alpha$ emission from ionized states of Fe 
in six NLS1s (\citealt{Comastriea98}, \citealt{TGN98}, 
\citealt{Vaughan99a}, \citealt{Comastriea01}, \citealt{TGN99}; 
\citealt{BIF01}).  
However, the observation of similar lines in some 
BLS1s, may point to the luminosity of the central source playing 
as important a role as accretion rate \citep[e.g.,][]{Guainazziea98}.

Alternative explanations of the extreme properties in NLS1s are that 
(1) the broad-line regions (BLR) of NLS1s have larger radii (i.e., the BLR gas 
is more distant from the nucleus) than in the BLS1s \citep{GFM83,MPJ96,WB98} 
then the narrow width of the lines is a reflection of the lower orbital velocity; 
(2) the NLS1s may be low-inclination (i.e., observed nearly face on) systems 
\citep{OP85}. Assuming the motions around the central source to be 
virialized, the narrowness of the lines is due to the fact that the
gas is moving preferentially on a plane that is almost perpendicular to
the line of sight, hence the line widths are reduced by a factor of 
$\sin i$, where $i = 0$ is face-on. 
One way to distinguish between these models is to measure the 
the size and virial mass of the BLR via reverberation techniques \citep{P93}; 
\citet{K2000} and \citet{P2000} find that the BLRs of NLS1s and BLS1s have 
comparable size, while NLS1s have virial masses one order of magnitude smaller 
than BLS1s, but cannot exclude the second model solely on the basis of 
reverberation results. 
However, \citet{BG92} and \citet{Joannaea00} disfavor the low-inclination model, 
while \citet{Nandraea97b} show that the inner regions of BLS1s also appear 
to be observed close to face-on.  


Tonantzintla (Ton) S180 (PHL 912) is a bright NLS1 with a 
low Galactic column density along the line-of-sight 
\citep[$N_{\rm H}=1.52 \times 10^{20}$ cm$^{-2}$;][]{Starkea92}. 
This source has FWHM H$\alpha$ and H$\beta \sim$ 900 km s$^{-1}$ 
and a redshift z=0.06198 \citep[][]{Wisotzkiea95}. Ton~S180 
has a relatively high luminosity, with absolute magnitude 
$M_{\rm B}=-23.1$ mag. 

Ton~S180 was observed by \ASCA{} on 1999 December 3 to 15, 
during a multi-wavelength  monitoring campaign that included 
observations from \HST, \RXTE, \Chandra, \EUVE, \FUSE{},  
in addition to optical--IR observations obtained from ground-based 
observatories. 
The results of the long-baseline timing project using 
\RXTE, \EUVE{} and \ASCA{} are reported in \citet{Edelsonea01}. The 
\HST{} and \FUSE{}  data were obtained contemporaneously with \Chandra{} and 
\ASCA{} and were undertaken to determine the spectral-energy-distribution 
of the source, as reported in \citet{Turnerea01c}. 
The \Chandra{} spectral result is reported in \citet{Turnerea01}. 
In this paper we present the results from the 
$\sim$ 400\,ks \ASCA{} observation of Ton~S180. 
This long observation has allowed us to study the variability of the 
spectral components on different timescales, from 12 days to $\sim 1$ day.
This is particularly interesting since the results of a 35-day long 
observational campaign on the NLS1 Arakelian (Akn) 564 indicate that
superimposed on a fast-varying continuum component that dominates the spectrum
is a slower-varying soft excess emission \citep{Akn564I,Edelsonea01}.
In \S\ref{dataobs} we describe our observations and data reduction. 
In \S\ref{timevar} we discuss the time variability of the source. 
In \S\ref{meansp} we analyze the mean spectrum  
and in \S\ref{ciccio} we discuss time-resolved spectroscopy. 
In \S\ref{results} we present a summary of our observational results,
and in \S\ref{comptonakn} a comparison with the properties of Akn~564.
Finally, in \S\ref{discussion} we discuss the  results.  
 
\vspace{0.8cm}

	\section{Observations and Data Reduction 
			\label{dataobs}}  

The focal-plane instruments on board \ASCA{} 
comprised two CCDs (the Solid-state Imaging 
Spectrometers SIS0 and SIS1, 0.4--10\,keV, \citealt{Burkeea91}) 
and two gas-scintillation proportional-counters (Gas Imaging 
Spectrometers  GIS2 and GIS3, 0.7--10\,keV, 
\citealt[][and references therein]{Ohashiea96});  these were 
operated simultaneously. 
\ASCA{} observed Ton~S180 for a total of $\sim$ 1 Ms, 
starting on JD 2451516.051 (for the screened data, see below).  
The data were reduced using standard techniques as 
used for  the {\it Tartarus}\footnote{ 
\anchor{http://tartarus.gsfc.nasa.gov}{http://tartarus.gsfc.nasa.gov}}  
database \citep{Turnerea99b}.  
Data screening yielded an effective exposure time of 
 338\,ks for SIS0,  368\,ks for the SIS1, and 
 405\,ks for both the GISs. 
The mean SIS0 count rate was 0.586 $\pm $ 0.001 \ctssec{} 
(0.5-10\,keV band). 
 
\label{SIStrouble} 
Increased dark current levels and decreased  
charge transfer efficiency (CTE) have become evident in the SIS 
detectors since $\sim 1993$, causing 
a divergence of SIS and GIS spectra. The degradation 
is not completely understood and at the time of writing not  
corrected for by any of the software (specifically, 
{\tt CORRECTRDD} does not correct for the effect). The instruments 
can diverge by as much as 40\% for energies $<0.6$\,keV for data 
taken in  2000 January\footnote{  
\anchor{http://heasarc.gsfc.nasa.gov/docs/asca/watchout.html} 
{see http://heasarc.gsfc.nasa.gov/docs/asca/watchout.html}}. 
The Ton~S180 data were calibrated using the latest calibration file
released on 2001 March 29 ($sisph2pi\_290301.fits$). 
The divergence of the SIS detectors at  
low energies can be compensated for in the spectral analysis.  
\citet{Yaqoob}\footnote{ 
\anchor{http://lheawww.gsfc.nasa.gov/$\sim$yaqoob/ccd/nhparam.html} 
{see http://lheawww.gsfc.nasa.gov/$\sim$yaqoob/ccd/nhparam.html}} 
provide a quantification of the apparent loss in SIS efficiency 
as a function of mission-elapsed 
time. 
The efficiency loss can be parameterized as a time-dependent  
absorption term (``excess \NH''). 
The correction for SIS0 follows a linear  relationship, 
$\NH(\mbox{SIS0}) = (\mbox{{\tt T}} -3.0174828 \times 10^{7}) 
 \times 3.635857508 \times 10^{12}$ cm$^{-2}$, where {\tt T} is the  
average of start and stop times of the observation, measured in   
seconds since launch. The SIS1 excess absorption term does not follow 
the simple linear form found for SIS0 but it is usually found that  
a slightly larger absorption column can be applied to the SIS1 data to 
bring it into line with SIS0.  For our observations, where 
{\tt T} $= 2.19 \times 10^{8}$ s, $\NH(\mbox{SIS0}) =6.9 \times 
10^{20}$ cm$^{-2}$ and we adopted $\NH(\mbox{SIS1}) = 1.0 \times 
10^{21}$ cm$^{-2}$. 
Application of the  excess \NH{} correction is important for this 
analysis, as it allows us to use the valuable data at energies  $<1$\,keV to 
examine the spectral variability of an interesting spectral component.

 	\section{The Time Variability\label{timevar}} 
 
Light curves were extracted using bin sizes of 256\,s and 
5760\,s  in the full-band (0.7-10\,keV)  for the SIS, the  soft-band 
(0.7--1.3\,keV) for the SIS data, and the hard-band (2--10\,keV)  
for both GIS and SIS data. 
We adopted 0.7\,keV as a lower limit for the SIS data, as appropriate for 
the setting of the SIS lower level discriminator for these observations. 
Light curves were constructed combining data from the SIS and GIS 
detector pairs. 
The exposure requirements for the combined light curves ensured 
the bins be fully exposed in each instrument for the 256\,s curves and at 
least 10\,\% exposed for the 5760\,s curves. 
The observed count rates correspond to a mean 2--10\,keV flux of 
$F_{\rm 2-10} = 6.5 \times 10^{-12}$ \eflux{} and 2--10\,keV luminosity 
$L_{\rm 2-10} = 4.9 \times 10^{43}$ \ergsec{}  
($H_0=75$ ${\rm km\ s ^{-1}\ Mpc^{-1},}$  $q_0=0.5$).  
	\centerline{\includegraphics[width=9.5cm,height=9.cm]{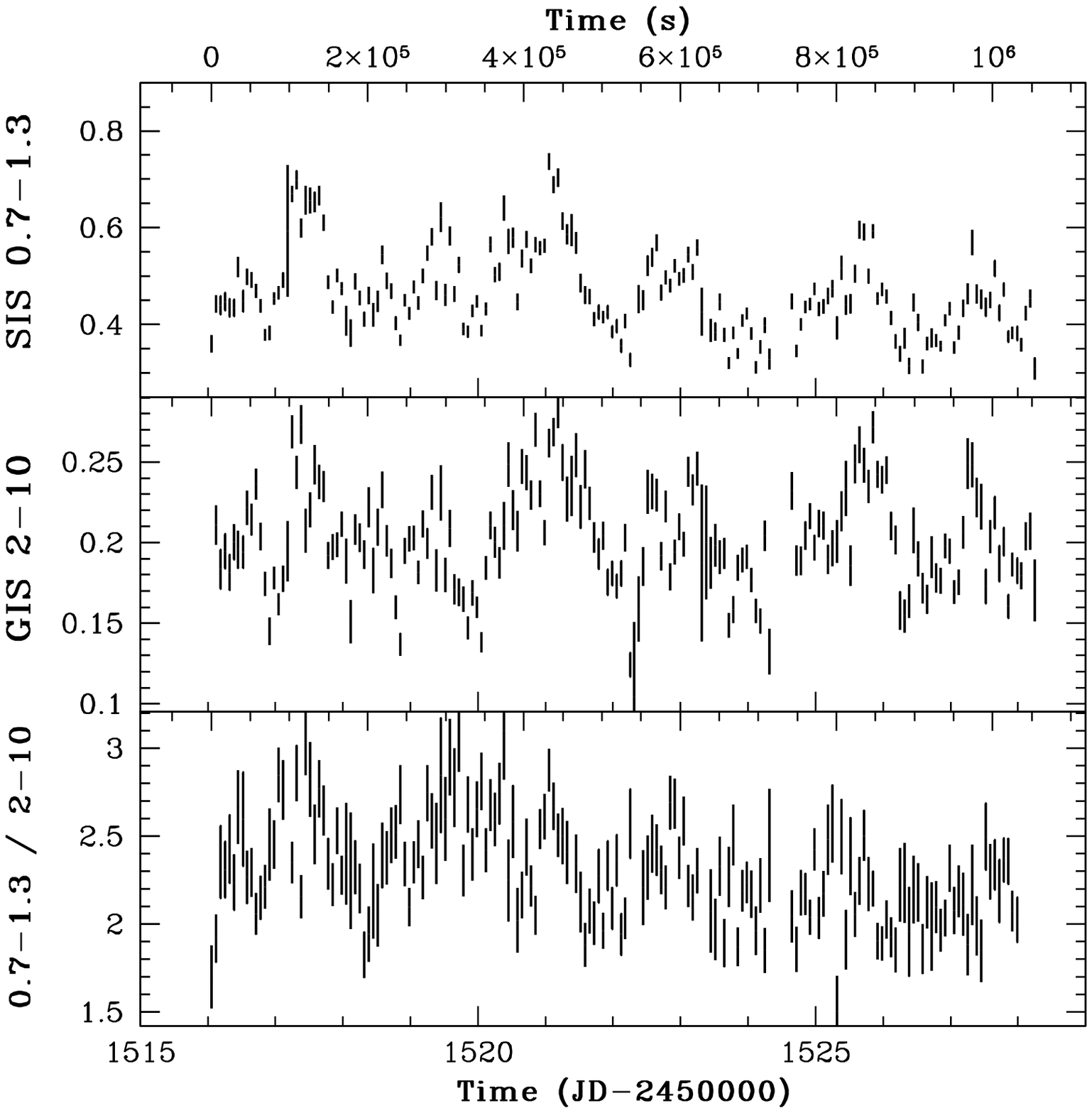}} 
        \figcaption{Light curves  
for the \ASCA{} data in \ctssec{} and in 5760\,s bins.  
The top panel is the SIS soft band (0.7--1.3\,keV)  
light curve; the middle panel the GIS hard band (2--10\,keV)  
and the bottom panel is the ratio of 0.7--1.3/2--10\,keV.  
The background level in the source cell is about 4\,\% and 10\,\% of the 
SIS and GIS source count rates, respectively, and not plotted. 
The times are reported both in seconds from the start of exposure (top axis) 
and in JD-2450000 (bottom axis). 
	\label{lcv2}} 
\vspace{0.5cm}
This mean flux level is $\sim$ 20\,\% brighter than that observed 
during a previous \ASCA{} observation on 1996 July 10 \citep[][]{TGN98}.
Figure~\ref{lcv2} shows the combined 0.7--1.3\,keV 
 SIS soft-band and  
GIS hard-band light curves in 5760\,s bins. 
The background levels in the source cells are about 4\% and 10\% of the 
SIS and GIS source count rates, respectively, and not plotted, or subtracted.
Figure~\ref{lcv2} also shows the softness ratio, defined as the ratio  
between the count rates in the 0.7--1.3 and 2--10\,keV bands.  
There is significant hardening of the spectrum during the observation,
and the softness ratio changes by $\sim 20$\,\%.   

The light curves integrated to 5760\,s show trough-to-peak flux variations 
by a factor of $\sim$ 2.7 in the 0.7--1.3\,keV band (SIS), 
$\sim$ 2.4 in the 2--10\,keV band (GIS).  
The light curves sampled on 256\,s reveal even larger  
amplitudes due to fast flickering, with a maximum amplitude of variability 
of a factor $\sim$ 3.5 for the SIS data in the 0.7--1.3\,keV band 
and $\sim$ 3.9 for the GIS data in the 2--10\,keV band. 
Figure~\ref{flare} shows the 0.7--1.3 
and 2--10\,keV light curves in 256\,s bins for three rapid ``events'' 
centered around 603, 608, and 614\,ks from the start of the observation  
(JD 2451523.03, 2451523.08, and 2451523.16). 
In the first event the soft and hard-band data show a 
variations (trough-to-peak) of factors of 
$R_{\rm max} {\rm (SIS)} = 1.32\pm0.13$ and 
$R_{\rm max} {\rm (GIS)} = 1.80\pm0.38$, respectively, 
in $\Delta t = 1972$\,s (the errors on $R_{\rm max}$ are obtained 
propagating the errors in the light-curve points). 
In the second event $R_{\rm max} {\rm (SIS)} = 1.34\pm0.13$ and 
$R_{\rm max} {\rm (GIS)} = 1.75\pm0.37$, respectively, 
in $\Delta t = 1972$\,s. 
In the third event, the variation is a factor of 
$R_{\rm max} {\rm (SIS)} = 1.33\pm0.13$ for the soft while 
the hard is consistent with a constant. 
These events correspond to a variation in luminosity of 
$\Delta L = 2.4 \pm 0.9 \times 10^{43}$ \ergsec{} (as calculated from SIS data,
from the first event; GIS data yield $\Delta L = 2.6 \pm 0.9 \times 10^{43}$ \ergsec). 
In the first and second events the hard X-ray flux variation is sharper 
than that in the soft band, and there is a    
change in the softness ratio, in the sense that when the flux 
increases the spectrum hardens, 
with a timescale of $\sim$ 1000\,s 
(Figure~\ref{flare}, bottom panel). 
However, in the third event a flux change is not accompanied by 
a strong spectral change.  Further examination of the light curves 
reveals other examples of diverse variability behavior with no obvious 
general trend. This is a behavior previously reported in \ROSAT{} 
observations  \citep[][]{Finkea97}. 

The soft and hard X-ray light curves in Figure~\ref{lcv2} exhibit 
similar characteristics, suggesting a short time delay 
between the variations in each curve. 
To quantify any correlations we undertook a cross-correlation 
analysis using the interpolation 
cross-correlation function (ICCF) method of \citet[][]{GasSpar86} 
and \citet[][]{GasPet87} as implemented by \citet[][]{WP94}. 
We calculated the CCFs of the total hard-flux in the 2--10\,keV band 
with respect to the total soft-flux in the 0.7--1.3\,keV band,
which has the highest signal-to-noise.  
The CCFs are sampled at a resolution of 0.05 day, and   
the centroids are computed using all points near the peak of the CCFs   
with values greater than 80\% of the maximum value of the 
correlation coefficient, $r_{\rm max}$. The 1-$\sigma$ uncertainties  
quoted for the ICCF centroid, $\tau_{\rm cent}$, are based on the 
model-independent Monte Carlo method described by \citet{Petersonea98}. 
We obtain  $r_{\rm max}=0.748$ and 
$\tau_{\rm cent}=0.025^{+0.024}_{-0.002}$ days.
Given the 180 points in our light curves, the probability 
of exceeding $r_{\rm max} \approx 0.3$ from uncorrelated samples is 
$\ll$ 0.1\,\%, and the lag is less than 0.07 d at 95\,\% confidence. 
Since the power-law component provides $\sim$ 72\,\% of the flux in 
the mean spectrum 0.7--1.3\,keV band, as we will show in \S\ref{meanhump}, 
and the power-law component dominates the 2--10\,keV band, 
this correlation is probably dominated by the rapid variability in the 
power-law flux. 
	\centerline{\includegraphics[width=9.5cm,height=8.5cm]{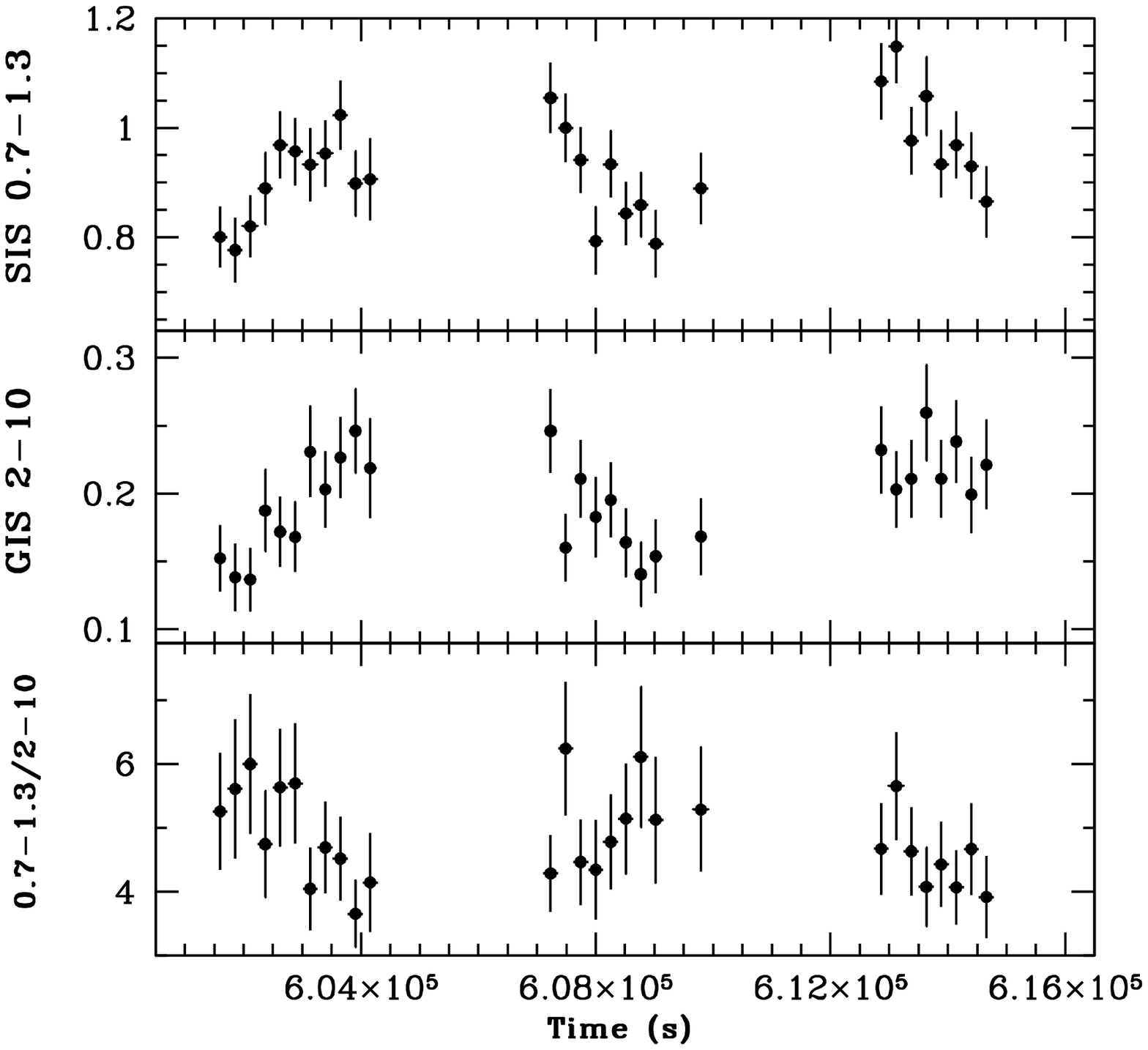}}
       \figcaption{
Same as Figure~\ref{lcv2}, for two ``events'' at JD $\approx$ 2451523. 
The time is seconds from the start of observation, the bin size is 256\,s.
In the first and second event, the SIS and GIS data show a variations 
up to a factor of 2 in $\Delta t = 1790$ s; 
in the third event, the variation is of a factor of $\sim$ 1.3 
in $\Delta t = 1536$ s (\S\ref{timevar}). 
	\label{flare}} 
	\centerline{}

\vspace{0.2cm}

	\centerline{\includegraphics[width=9.5cm,height=8.5cm]{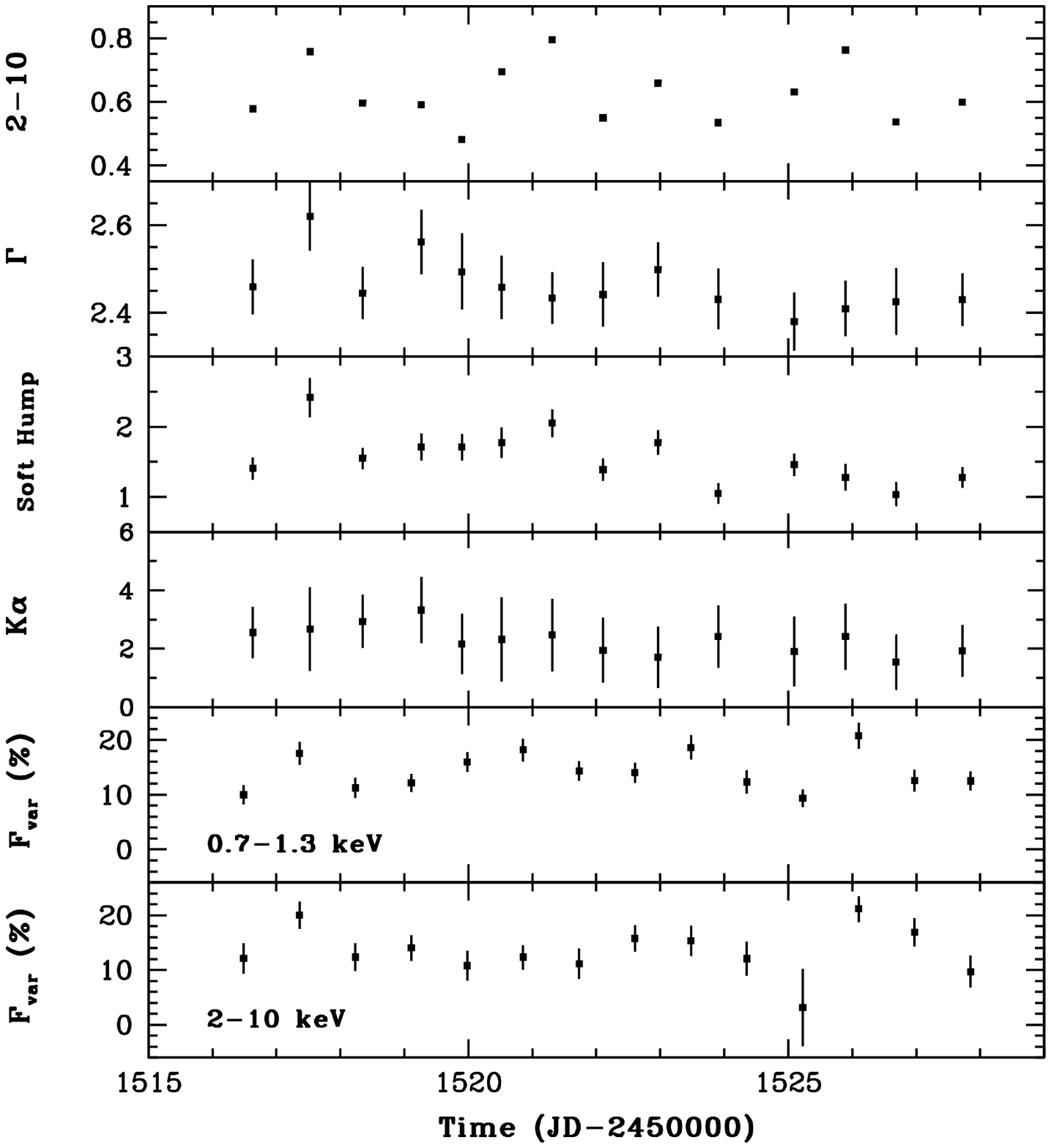}}
	\figcaption[Spectral Parameters]{Spectral and timing parameters  
obtained from fits to the individual time-resolved spectra. 
From the top, the light curves are  
the (model) continuum flux in the hard band,  
the photon index $\Gamma$,  
the soft hump flux, 
the $\Kalpha${} flux ({\tt laor}),  
the fractional variability $F_{\rm var}$.  
The continuum fluxes are in units of 
$10^{-11}$ \eflux, the soft hump flux in units of $10^{-2}$ \phflux{}
and the $\Kalpha${} flux in units of $10^{-5}$ \phflux. 
	\label{14fits}} 
	\centerline{}

	\subsection{Fractional Variability Amplitude\label{fvar}} 

The fractional variability amplitude $F_{\rm var}$ and its error 
$\sigma_{F_{\rm var}}$ are  defined  in \citet{Edelsonea01}, as  
\begin{equation} 
F_{\rm var} = \sqrt{\frac{S^2 - \langle \sigma^2_{\rm err}  
\rangle}{\langle X \rangle^2} } , \;\;\;\;\;\;\;\;\;\;\;
\sigma_{F_{\rm var}} = \frac{1}{F_{\rm var}}  
		\sqrt{\frac{1}{2N}} \frac{S^2}{\langle X \rangle^2} .
\end{equation} 
where $S^2$ is the total variance of the light curve,  
$\langle \sigma^2_{\rm err} \rangle$ is the mean squared error, and  
$\langle X \rangle$ is the mean count rate.  
 
First we calculated $F_{\rm var}$ across the baseline 
of the entire observation.  This quantity measures deviations 
compared to the mean, integrated over the entire 12 days.
We measured this quantity in the soft (0.7--1.3\,keV)  
and hard (2--10\,keV) bands, using our light curves with 256\,s bins. 
These bands were chosen to sample energy ranges 
containing different spectral components (see \S\ref{meansp}) 
while still maintaining good signal-to-noise. 
$F_{\rm var}$ thus calculated is 19.12 $\pm$ 0.58\,\% in the 
0.7--1.3\,keV band, and 17.26 $\pm$ 0.65\,\% in the hard-band, 
using SIS data for both tests. 

To examine the evolution of timing properties of Ton~S180 further, 
we split the data into 14 evenly-sampled sections across the 12 
day \ASCA{} observation. Even sampling is important because 
$F_{\rm var}$ depends strongly on the duration of the data-train.  
The baseline for each time-selected section of data was 75\,ks 
with average on-source exposure time 25\,ks.
We then calculated $F_{\rm var}$ over each of these one-day intervals 
(again with 256\,s bins in each ``daily'' light curve''). 
This test showed that $F_{\rm var}$ is significantly 
variable showing correlated changes in the soft and hard bands, 
on a day-to-day basis (figure~\ref{14fits}). 

As we will demonstrate in \S\ref{meanhump},  
in the mean spectrum the power-law continuum provides about 72\,\% of 
the flux in the 0.7--1.3\,keV band, and the power-law variations 
contribute  significantly to $F_{\rm var}$ in both the soft and hard bands,
explaining the gross similarity between the quantity in those two bands.  
The 

\vspace{-1.0cm}

\centerline{\includegraphics[width=10.5cm,height=10.0cm]{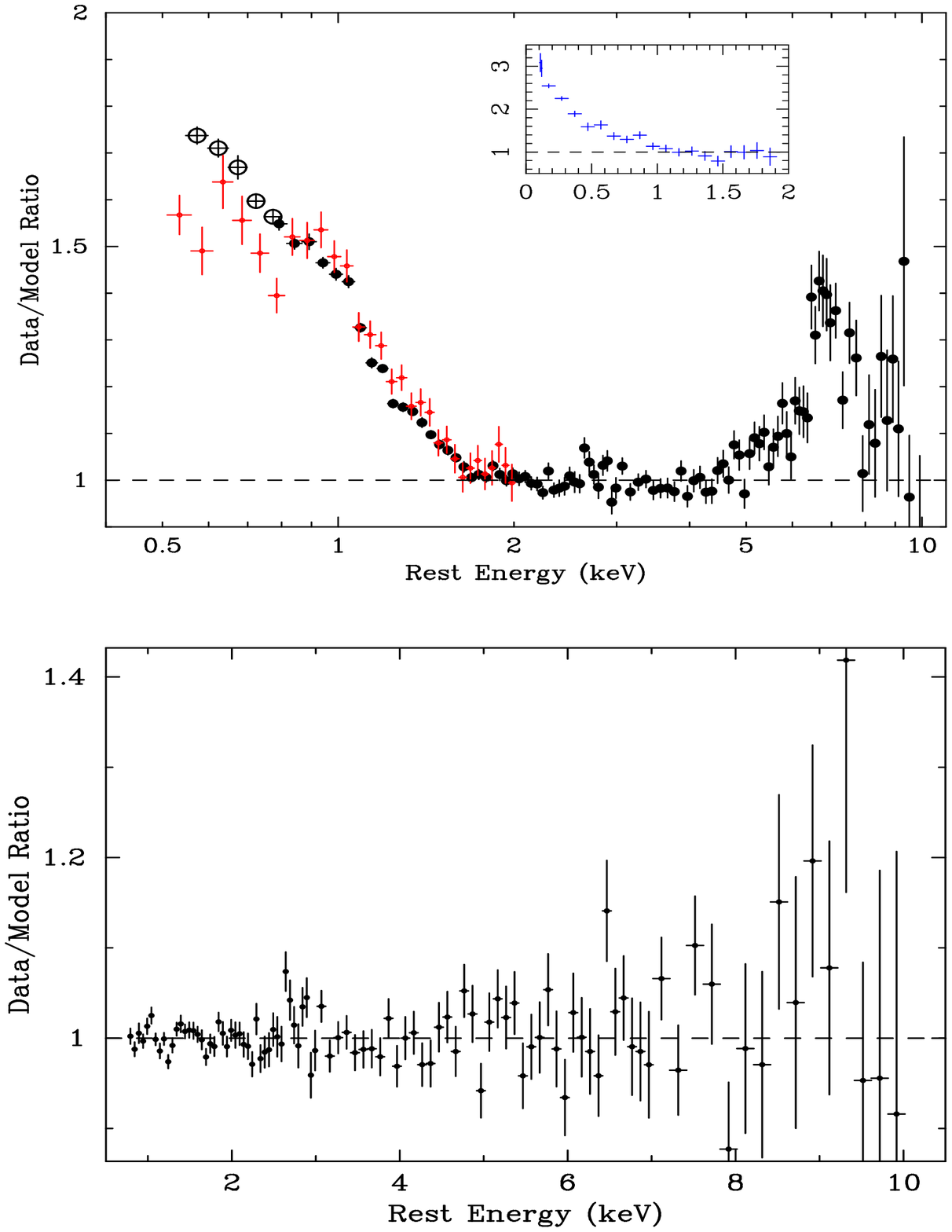}}

\vspace{-1.0cm}

\figcaption{Top: Data/Model ratio where the model  
is a simple power law fit to the 1.8--4.71\,keV data  
(observer's frame). The rest of the combined SIS and GIS data are then overlaid,  
revealing a strong soft hump and Fe emission line, as filled circles.
The 0.5--0.7\,keV SIS data, though not used, are also plotted as 
empty circles. The crosses are the \ASCA{} data from 1996 below 2\,keV.
The inset panel shows the {\it ROSAT} PSPC data compared to the 
continuum power law, illustrating the shape of the soft hump at 
lower energies (\S\ref{meansp}). 
All datasets confirm the presence of the soft hump, and detail 
its shape. 
Bottom: Data/Model ratio where the model is a power law plus 
soft hump (Gaussian; \S\ref{meanhump}) plus 2-Gaussian model 
for the Fe K$\alpha$ line (\S\ref{meanfek}).
\label{pl_rat}} 
\centerline{}
\vspace{0.3cm}

\noindent
slightly greater $F_{\rm var}$ for the soft band is probably due to 
a combination of changes in the relative flux levels of 
two spectral components (power law and hump), and slope changes in 
the power-law, as we will show in \S\ref{humpfits}.

	\section{The Mean Spectrum\label{meansp}} 
 
Source counts were binned  with a 
minimum of 20 counts per energy bin for the spectral analysis. 
The data from the four instruments were fit simultaneously, 
with the relative normalizations free to allow for small 
differences in calibration  
of the absolute flux and in the fraction of encircled 
counts encompassed by the SIS versus GIS extraction cells.  
Spectral fits were performed using {\tt XSPEC V11.0.1}  
and response files generated with {\tt HEAsoft v5.0.4}.
Ton~S180 is detected at better than 3-$\sigma$ level in the 
SIS data for energies $\lesssim 7.5$\,keV 
and in the GIS data for energies $\lesssim 9.5$\,keV
(observed frame, in the unbinned spectrum).

The photon index $\Gamma$ (defined by photon flux $P_E \propto
E^{-\Gamma}$) was determined by fitting 
a power-law model attenuated by Galactic absorption  
with an additional correction applied to SIS0 and SIS1 
to compensate for the low-energy degradation as described  
in \S\ref{SIStrouble}.  
For this fit we used data in the bandpass 1.8--4.71\,keV  
(observer's frame). 
The power-law fit yielded $\Gamma = 2.44 \pm 0.02$ and  
$\chi^2= 720$ for 683 degrees of freedom ($dof$). 
Inclusion of the data in the  7.06--9.42\,keV bandpass produces 
consistent results.  This and subsequent errors represent, 
unless otherwise specified, the 90\,\% confidence level. 

The data/model ratio is shown in Figure~\ref{pl_rat}, top panel. 
A strong soft excess is evident, appearing as a hump of emission   
rising above the power-law continuum at energies $<2$\,keV. 
This feature was also observed in the \ROSAT{} PSPC observations 
\citep[][inset in top panel of Figure~\ref{pl_rat}]{Finkea97},
\BeppoSAX{} observations \citep{Comastriea98}, 
and the earlier \ASCA{} observation 
\citep[sequence number 74081000;][]{TGN98}. 
Hereafter we refer to this component as the ``soft hump''.  
Also evident in the main panel is an excess of emission close to 
7\,keV, which is due to an unmodeled  Fe K$\alpha$ emission line. 

Minor calibration problems are visible in such 
high signal-to-noise data at $\approx$ 1.7--3.0\,keV. 
They are not so serious as to warrant exclusion 
from the fit. We note, however, that they contribute significantly 
to the $\chi^2$ in the soft component fits (\S\ref{meanhump}) and to a lesser 
extent to the K$\alpha$ fits (\S\ref{meanfek}).

	\subsection{The Soft Component\label{meanhump}}

We confirm the presence of the soft X-ray emission component 
previously observed by \citet{Finkea97}, \citet{Comastriea98},  
and \citet{TGN98}. At the time of writing, the status of the SIS 
calibration limits the accuracy to which the absolute form of 
the soft spectral hump can be determined. Thus we adopted a 
simple parameterization of the soft hump which allowed us to 
perform a sensitive examination of the flux variability of the 
component. 
The simultaneous \Chandra{} LETG data for Ton~S180 \citep{Turnerea01} 
and the results for NGC~4051 \citep{Collea01} show that the soft hump 
is a smooth continuum component, as opposed to a blend of unresolved 
spectral features. 
The variability study afforded by the \ASCA{} data is therefore 
the optimum available tool for exploring the nature of the soft hump. 
This was also shown to be the case for Akn~564 \citep[][]{Akn564I}.  

We used SIS data in the range 0.7--4.71\,keV simultaneously with 
GIS data in the range 1.0--4.71\,keV (observer's frame). The lower energy
limit of the SIS data was based on the level of agreement achieved 
between the two CCDs using our methods of correction (\S\ref{SIStrouble}). 
The upper  energy limit of both SIS and GIS effectively excluded the 
Fe K$\alpha$ regime for these fits, in order to avoid an overly complex 
model (which can result in false or local minima being found).
A steep power-law component is an inadequate representation of this 
excess, as its form shows some curvature. 
The Gaussian model which best fits the soft hump   
has an energy of the peak $E=0.17^{+0.17}_{-0.17p}$\,keV 
(the energy pegged at the lower limit), full width at half maximum
FWHM $ =1.01^{+0.05}_{-0.12}$\,keV  
and flux $1.50^{+0.37}_{-0.61} \times 10^{-2}$ $\phflux$ 
corresponding to a mean equivalent width EW $=94^{+23}_{-38}$\,eV.  
For this fit $\chi^2=1069$ for 896 $dof$, but there is a contribution to 
the $\chi^2$ from the $\approx$ 2.5--3.0\,keV region due to calibration 
problems (see \S\ref{meansp}). 
Significantly inferior (to better than 99.9\,\% confidence; 
$\chi^2=1096$ for 897 $dof$) is a parameterization of the soft hump as a 
blackbody model; the best fit that yielded a rest-frame temperature 
$kT=153^{+2}_{-3}$\,eV, 
flux $1.07^{+0.03}_{-0.04} \times 10^{-4}$ $\phflux$,
and absorption corrected luminosity in the 0.7--1.3\,keV band 
$L_{\rm 0.7-1.3}= 1.6 \times 10^{43}$ \ergsec. 

Using a Gaussian parameterization of the soft hump, we  
determined that the power-law continuum contributes to $\sim 72$\,\% 
of the flux in the soft (0.7--1.3\,keV) band of the mean spectrum. 
We also retrieved the 1996 ASCA spectrum \citep{TGN98} from the 
{\it Tartarus} database \citep{Turnerea99b}, and found consistent
contributions of the different spectral components to the soft band  
over the same energy ranges as we used for our 12-day observation. 

\vspace{1.0cm}

\footnotesize   
\begin{center} 
{\sc TABLE 2\\   
Spectral Fits in the 2--10\,keV Band.\label{kafitstab}}
\vskip 4pt
\begin{tabular}{lcccc}
\hline
\hline  
{Model} & {$\Gamma_{2-10}^{a}$} 
           & {$E_{\rm Fe}^{a,b}$} & {EW$^{a}$} 
	& {$\chi^2/dof$}\\
      {} & {} & {(keV)} & {(eV)} & {} \\
	  {(1)} & {(2)} & {(3)} & {(4)} & {(5)} \\
\hline
 PL$^{c}$           & $2.37 \pm 0.01$ & ... & ... & 1518/1372 \\
 PL + {\tt DISKLINE}      & $2.43^{+0.01}_{-0.02}$ & $6.40^{+0.27}_{-0.00p}$  & $461^{+120}_{-84}$ & 1353/1368 \\
 PL + {\tt LAOR}    & $2.43 \pm 0.02$ & $6.55^{+0.16}_{-0.15p}$ & $517^{+123}_{-111}$ & 1352/1368  \\
 PL + NG      & $2.39 \pm 0.01$ & $6.75^{+0.08}_{-0.04}$  & $182^{+30}_{-26}$ & 1404/1370 \\
 PL + BG      & $2.44 \pm 0.02$ & $6.71^{+0.12}_{-0.14}$  & $550^{+184}_{-128}$ & 1352/1369 \\
 PL + BG + NG & $2.46^{+0.03}_{-0.02}$ & $6.58^{+0.25}_{-0.28}$ & $512^{+225}_{-136}$ &  1328/1359 \\
	      &                        & $6.81^{+0.08}_{-0.12}$ & $90^{+33}_{-31}$    &  \\ 
\hline
\end{tabular}
\end{center}
$^{a}$ 90\,\% confidence level uncertainties. \\
$^{b}$ Rest frame of Ton~S180. \\ 
$^{c}$ Over the whole hard energy range. The value of the photon index 
from the continuum fit is $2.44 \pm 0.02$ (\S\ref{meansp}). \\
\setcounter{table}{1}
\normalsize
\centerline{}

	\subsection{The Fe K$\alpha$ Regime\label{meanfek}}  

\citet{Comastriea98} and \citet{TGN98} found evidence for line 
emission at $\sim$ 7\,keV, indicative of an origin from material 
containing ionized iron, consistent with the iron  
line from H-like iron at 6.94\,keV. Hence, having found an adequate 
parameterization of the continuum shape (in the 1.8--4.71\,keV band), 
we included the 4.71--7.45\,keV SIS data and the 4.71--9.42\,keV GIS data
in the analysis and examined the data/model ratio (versus the best-fit 
continuum model, top panel of Figure~\ref{pl_rat}). 
The mean line profile is broad and asymmetric, similar to that 
observed in the previous \ASCA{} observation \citep[][]{TGN98}. 

We utilized the SIS data in the 
1.8--7.45\,keV range simultaneously with the GIS data in the 
1.8--9.42\,keV range. Table~\ref{kafitstab} summarizes our results: 
Column (1) lists the models, Column (2) the photon index, 
Column (3) the rest-frame energy of the fitted Fe line, 
Column (4) its equivalent width, and Column (5) $\chi^2$ 
and degrees of freedom relative to the fit. 
PL indicates the power-law continuum, BG the broad Gaussian,
and NG the narrow Gaussian (see details below).

The asymmetry of the profile prompted us to fit the iron K$\alpha$ 
line using the {\tt diskline} model profile of \citet{Fabianea89}.
This model assumes a Schwarzschild metric, with an emissivity law 
$r^{-q}$ for the illumination pattern of the accretion disk, 
where $r$ is the radial distance from the black hole.
We adopted $q=2.5$ based upon the results of \citet{Nandraea97b}. 
We also assume that the line emission originates within 1000 gravitational 
radii ($R_{\rm g} = G M / c^{2}$). 
The inclination of the system is defined such that $i=0$ is a disk  
oriented face-on to the observer. 
Fits using this model yielded $\chi^2=1353$ for 1368 $dof$. 
The rest-energy of the line was $E = 6.40^{+0.27}_{-0.00p}$\,keV,   
the inner radius was $r=6^{+10}_{-6p}\ R_{\rm g}$,  
the inclination was $i=35^{+22}_{-35p}$ degrees, 
and flux $2.06^{+0.54}_{-0.37} \times 10^{-5}$ \phflux. 
The equivalent width was EW $=461^{+120}_{-84}$\,eV.  

We also tested a model for the line profile assuming a Kerr metric
for a maximally rotating black hole as implemented by \citet{Laor91}. 
This will have the most intense gravitational effects.  
We fixed the emissivity index as for the Schwarzschild case and 
the outer radius at the maximum value allowed by the model, 400 $R_{\rm g}$.
Using the same energy restriction as for the Schwarzschild model, 
the Kerr model provides a fit-statistic $\chi^2 = 1352$ for 1368 $dof$.
The rest-energy of the line was $E = 6.55^{+0.16}_{-0.15p}$\,keV, 
inclination was $i=23^{+14}_{-23p}$ degrees, and 
flux $2.22^{+0.53}_{-0.47} \times 10^{-5}$ \phflux. 
The equivalent width was EW $=517^{+123}_{-111}$\,eV.

\centerline{\includegraphics[width=9.5cm,height=11.5cm]{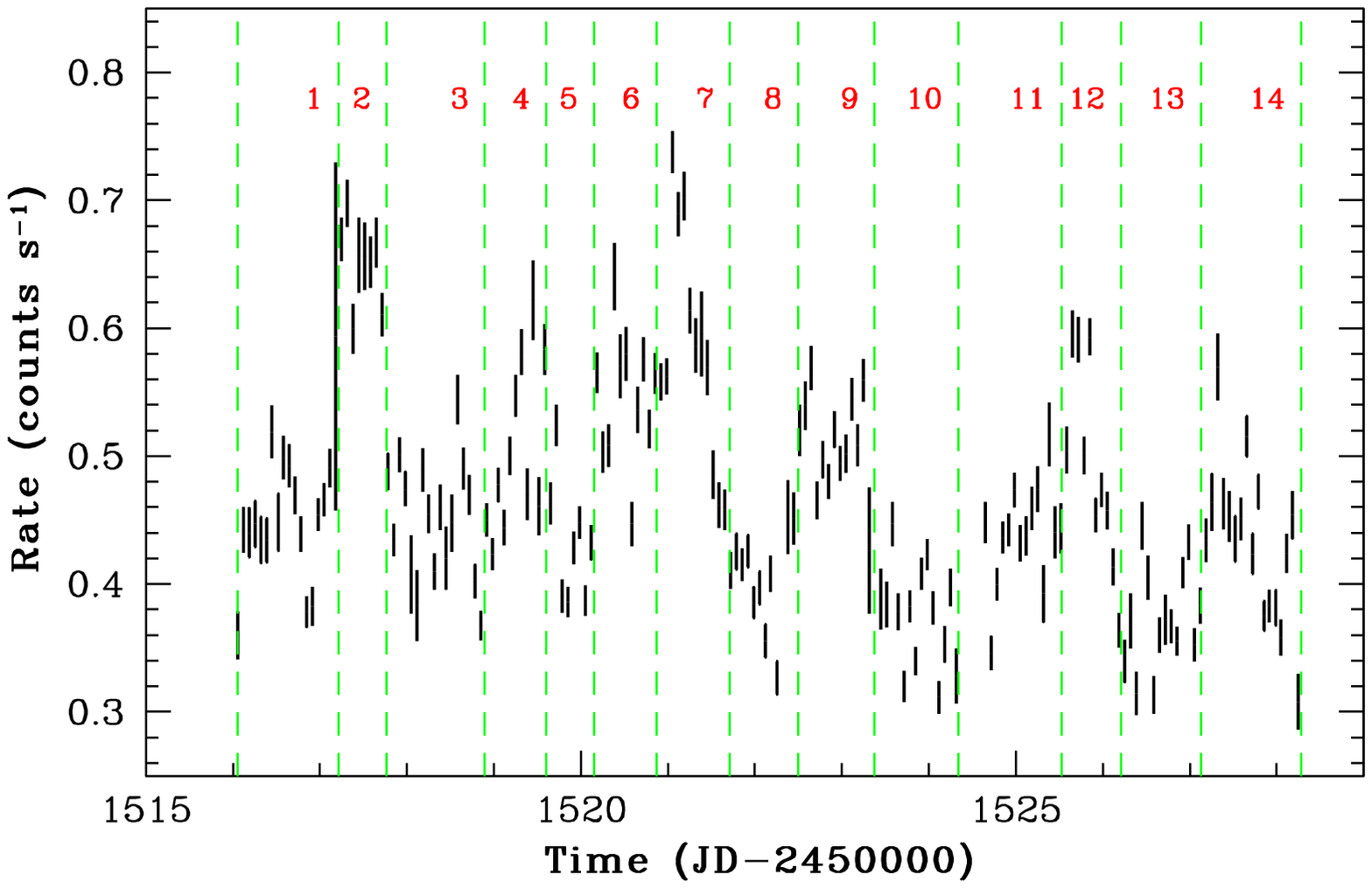}}
\vspace{-5cm}
\figcaption[SIS Light curves]{Combined SIS 0.7--1.3\,keV light  
curve in \ctssec{} and in 5760\,s bins. The background level in  
the source cell is about 4\,\% of the source count rate, and not  
plotted or subtracted. The vertical dashed lines show our 14 time intervals 
within which spectra were extracted (\S\ref{ciccio}). 
The ``flares'' shown in Figure~\ref{flare} (\S\ref{timevar}) 
are within bin 9. 
\label{lcvcuts}} 
\centerline{}
\vspace{0.5cm}

The line was better fit as the sum of a broad and a 
narrow redshifted Gaussian profile, with $\chi^2=1328$ for 1359 $dof$ 
(a broad or narrow Gaussian alone are a significantly 
worse fit at $>$ 99\,\% confidence; see Table~\ref{kafitstab}). 
The rest energy of the narrow line (fixed at 10\,eV width) 
was $E_{\rm N}=6.81^{+0.08}_{-0.12}$\,keV, with EW $ = 90^{+33}_{-31}$\,eV. 
The broad component gave $E_{\rm B}=6.58^{+0.25}_{-0.28}$\,keV and 
EW $ = 512^{+225}_{-136}$\,eV.  
The residuals of the 2-Gaussian fit are shown in the bottom panel of 
Figure~\ref{pl_rat}.
Some excess emission is evident in the 8--9\,keV region.
However, the signal-to-noise in this energy range is so low
(the detection level is $\sim 3$\,$\sigma$) that 
no meaningful spectral fit can be performed. 

We do not detect any narrow (fixed at 10\,eV width) \Kalpha{} emission 
at rest energy 6.4\,keV, in addition to the Schwarzschild, Kerr, single 
broad Gaussian, or 2-Gaussian models.

\section{Spectral Variability\label{ciccio}}	 
	\subsection{Method and Selection Details\label{cutmethod}}  

To examine the spectral evolution of Ton~S180, we created 14 time-selected 
spectra across the 12 day \ASCA{} observation by sampling throughout 
the light curve following flares and dips with {\tt Xselect V2.0}.  
The resulting average baseline for each time-selected spectrum was
75\,ks, for an average on-source exposure time 25\,ks.
Our choice of intervals is shown in Figure~\ref{lcvcuts}.
The ``events'' described in \S\ref{timevar} (Figure~\ref{flare})  
occur during time bin 9. 
Background, ancillary response, and response matrix files were set to be
those of the mean spectrum. The background spectrum and flux 
did not vary significantly during the observation and use of the mean 
spectra yielded the best possible signal-to-noise for the time-resolved 
spectroscopy. 
Again, spectra were binned to achieve a minimum of 20 counts per energy bin. 
We performed all fits by fixing the scaling factors for instrument 
normalization based upon the fits to the mean spectrum. 
Corrections for the SIS low-energy problem were fixed at the same 
values used for the mean spectrum. 
All models were modified with a Galactic absorption. 
Time assignments for the spectra refer to the mid-point of the observation, 
in JD $-$ 2450000. The results of our analysis are shown in Figure~\ref{14fits} 
in the form of time series curves for the various parameters. 
These are described in detail in the following sections.

\vspace{-2.5cm}

\centerline{\includegraphics[width=10.cm,height=12cm]{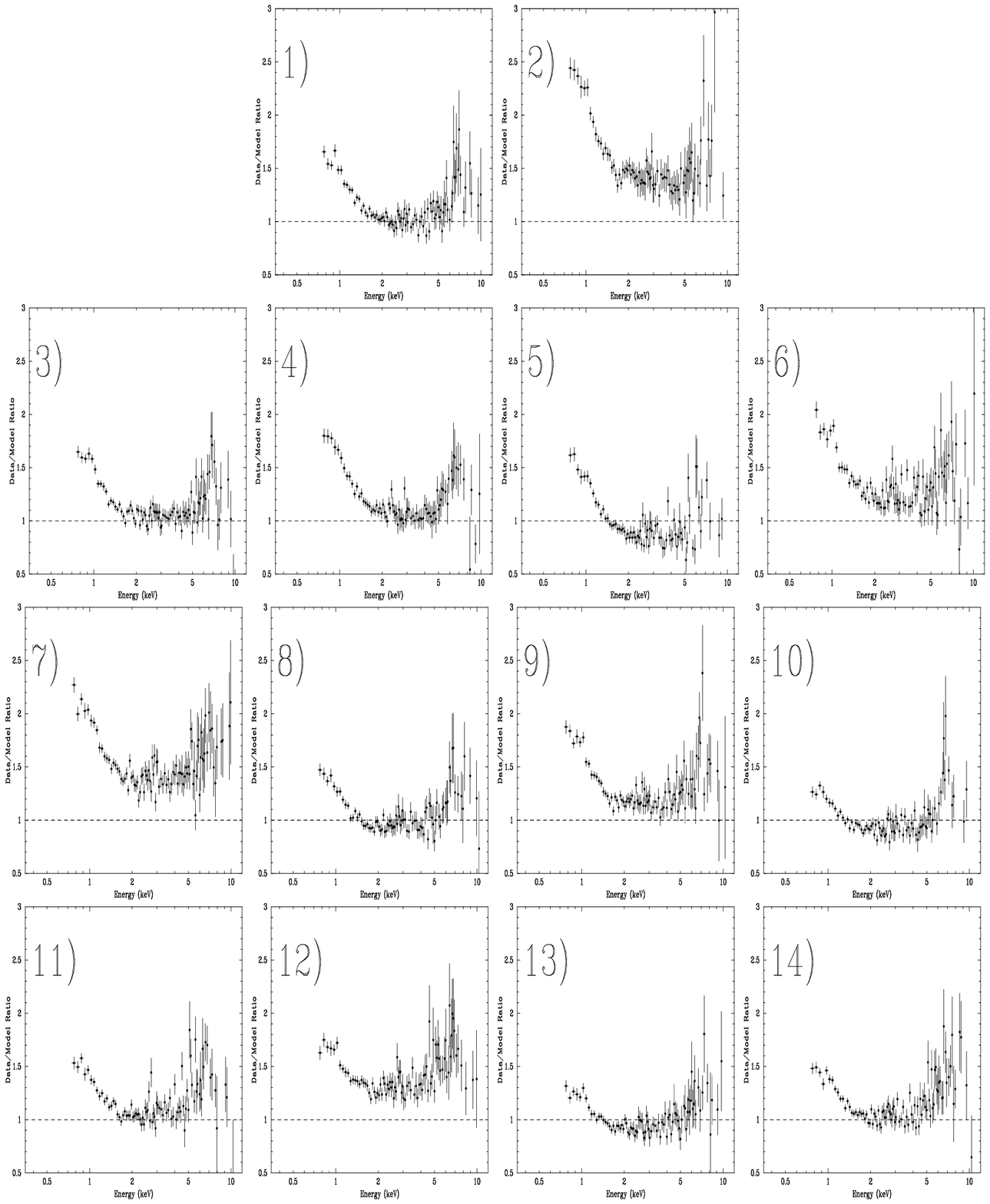}}
\vspace{-0.5cm}
\figcaption[Ratio plots]{Ratio plots obtained by fitting the 
best-fit model for the first spectrum to the following 13 spectra.
Energy ranges and instrument utilized are described in 
\S\ref{contfits}. Energies are in the rest frame of Ton~S180. 
\label{allratios}}
\centerline{}

	\subsection{Variability of the Continuum\label{contfits}} 

We fit each of the 14 time-selected spectra in the 1.8--4.71\,keV range 
(both SIS and GIS; observer's frame) with a simple power-law model.  
The light curves for the (model) continuum flux and the best-fit 
photon index $\Gamma$ are shown in Figure~\ref{14fits}.
The photon index $\Gamma$, thus sampled on a $\sim 1$\,day timescale,  
ranges in 2.38--2.62 (i.e., $\Delta\Gamma=0.24$) 
across the 12 days, but these variations are not significant
(fitting the photon indices to a constant model yields 
$\chi^2=10$ for 13 $dof$). 
Therefore, the rapid flux variability evident on timescales as short as 
$\sim 1000$\,s in the hard band (Figure~\ref{flare}, \S\ref{timevar}) 
is not due to changes in $\Gamma$, but is truely flux changes in that band. 
However, on longer timescales of about a week significant changes in $\Gamma$ occur, 
as we will show in \S\ref{humpfits}. 

Figure~\ref{allratios} shows ratio plots obtained by comparing the 
best-fit model for the first spectrum to the following 13 spectra, without  
performing any fit. The plot illustrates the variations of the spectral shape  
and flux compared to the first day of data.   
Strong variations in the flux and, to a lesser degree, in the continuum slope
are evident.  
The soft hump is always present superimposed on  the power-law continuum, 
although it appears to change in absolute strength, relative to the 
power-law continuum and possibly in shape (see, for example 
time-cuts 2 and 13) on a timescale of approximately a day. 
Variations in the shape of the soft hump, however,
are a minor effect compared to the flux variations of the soft hump.

\centerline{\includegraphics[width=12.5cm,height=14cm]{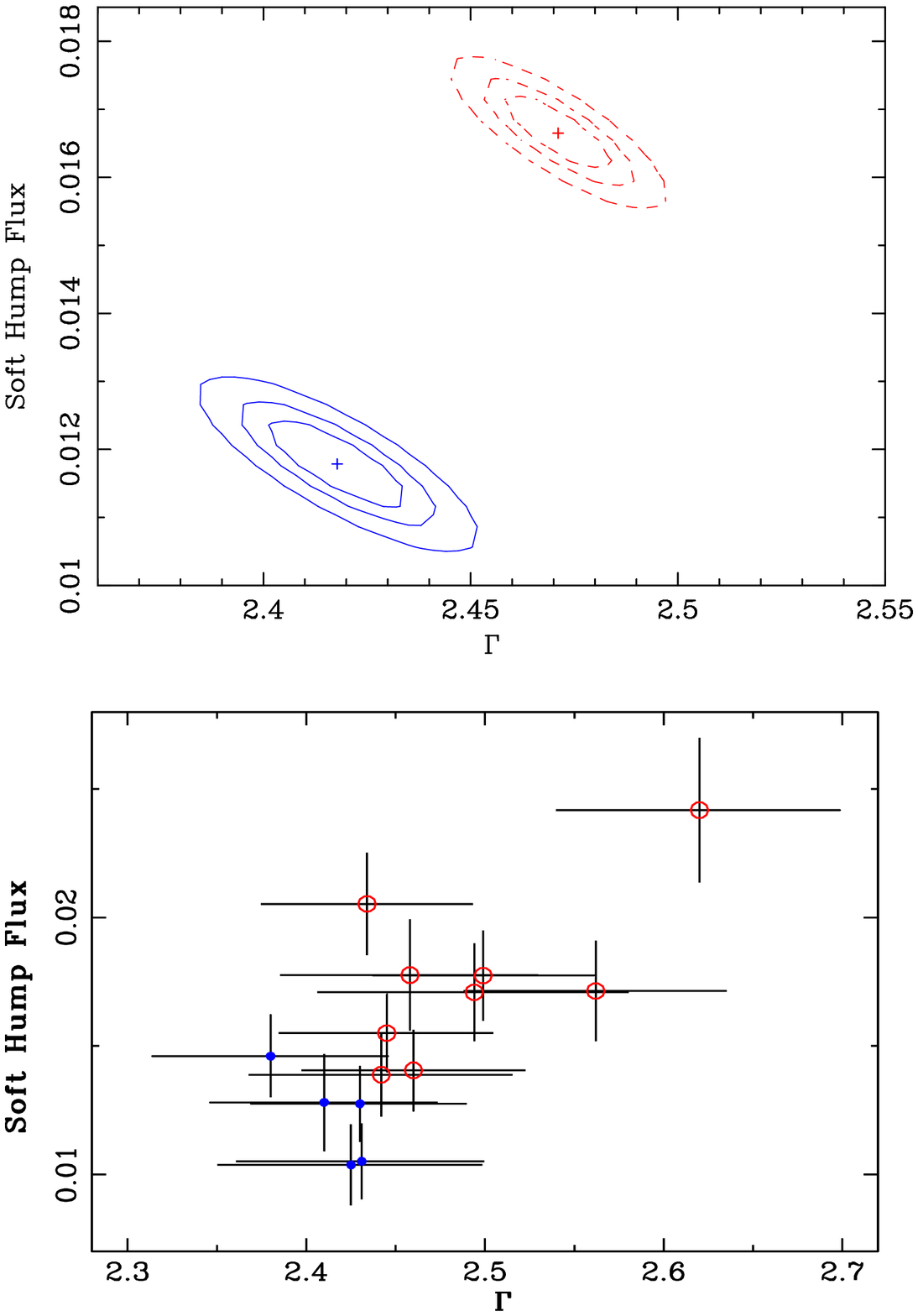}}
\vspace{-1.5cm}
\figcaption[Soft Hump versus Photon Index]{ 
Top panel: the  $\Delta \chi^2 = 2.3, 4.61, 9.21$ contour levels for 
the soft hump flux (in units of \phflux) vs.\ photon index $\Gamma$. 
The full contours correspond to the hard state (\S\ref{kafits}),
the dashed contours to the soft state, 
and crosses indicate the best-fit values.
Bottom panel: the strength of the soft hump plotted against photon 
index $\Gamma$ in our 14 time-selected spectra 
with high-state (open circles) and low-state (filled circles) points 
overlaid.
\label{gamhump}} 
\centerline{}

	\subsection{Variability of the Soft X-ray Hump \label{humpfits}}

We examined the variability of the soft hump using SIS data in the range 
0.7--4.71\,keV simultaneously with GIS data in the 1.0--4.71\,keV range 
(observer's frame), as we did for the mean spectrum in \S\ref{meanhump}.  
The model was a simple power-law plus a broad Gaussian 
for the soft hump, with the Gaussian peak and width fixed 
at the values noted in \S\ref{meanhump}. The flux of the 
soft hump, the flux of the continuum power law, and the 
power-law slope were free parameters. 
The time series for the flux of the soft hump is shown in 
Figure~\ref{14fits}. Fitting a constant model to this flux yields 
$\chi^2=47$ for 13 $dof$. The soft hump flux varies by a factor 
of $R_{\rm max} = 2.33 \pm 0.45$ while the 2--10\,keV flux 
(when binned the same way) varies by a factor of 
$R_{\rm max} = 1.65 \pm 0.02$. 

In Akn 564, \citet[][]{Akn564I} showed that the different amplitudes 
of variability of the soft hump and power-law 
($6.44\pm3.30$, compared to $3.97\pm0.06$) 
explain the gross change in softness ratio across the observation.
In Ton S180, the amplitude of variation of the soft hump is not so pronounced,
although it is clear from the light curves (Figure~\ref{14fits}) that 
most of the flares in the hard X-ray flux are also present in the 
soft hump. 
To clarify the situation in Ton S180, we split the data into 
``hard'' and ``soft''  states based upon the softness ratio. 
We chose {\tt T} $< 6.3 \times 10^{5}$ s for our soft-state spectrum, 
and {\tt T} $ > 6.3 \times 10^{5}$ s for our hard-state spectrum 
where {\tt T} is the time from the start of the observation 
(see Figure~\ref{lcv2}).
Fits were performed on each state, as we did for the mean spectrum 
(\S\ref{meanhump}), and confidence contour plots were generated for the 
flux of the soft hump versus the photon index,  
assuming the soft hump shape to be approximately constant during 
the observation. 
The resulting $\Gamma$--soft hump flux contours are well separated
(Figure~\ref{gamhump}, top) in both directions. Hence,  
the confidence contour plot indicates that the source 
varies significantly in photon index {\it and} strength of the soft hump on a 
timescale of about one week. 
Therefore both effects are contributing to the change in softness ratio 
observed across the baseline of the observation. The spectral changes 
observed between the soft and hard states result in a change 
of the EW of the soft hump from $EW_{\rm soft}=100\pm5$\,eV to 
$EW_{\rm hard}=84^{+6}_{-7}$\,eV. 

Since we adopt a fixed soft hump shape in our fits of the 14 time-selected 
spectra, we are confident our fitting method can separate the fluxes 
of soft hump and power-law continuum. 
Furthermore, the bottom panel in Figure~\ref{gamhump} shows $\Gamma$ plotted 
against the strength of the soft hump component and demonstrates a 
strong correlation (Spearman rank coefficient $r_{\rm s} = 0.68$, 
probability of chance occurrence is 8$\times 10^{-3}$), 
as is expected from disk-corona models, as we will see in \S\ref{discussion}. 
If the strength of the soft hump component and the photon index were difficult
to separate in the spectral fit, we would expect an anticorrelation instead.
Figure~\ref{gamhump}  also shows that a steeper photon index appears 
to be associated with a relatively strong soft hump. 
Finally, as noted in \S\ref{contfits}, the variations in the shape of the 
Gaussian soft hump are small so that our model for the soft hump parameterizes 
most of the flux in each time-selected spectrum, and does not weaken our approach 
in fixing the shape when testing for flux variability.

	\subsection{Variability of the Fe Emission Line\label{kafits}} 

As we have seen in \S\ref{contfits}, Figure~\ref{allratios} suggests that
there are  small variations in the strength of the Fe emission line.   
To investigate this possibility, we utilized SIS data in the 
1.8--7.45\,keV range simultaneously with the GIS data in the  
1.8--9.42\,keV range, as we did for the mean spectrum in \S\ref{meanfek}.
The model was a simple power-law continuum plus different models for the 
K$\alpha$ line, with fixed shape parameters. 
We initially tested the assumption of a fixed shape 
by splitting the data with an intensity division equivalent to an SIS0 
count rate of 0.6 \ctssec. This yielded high- and low-state spectra with 
similar signal-to-noise.  
We also calculated confidence contours of line flux versus $\Gamma$. 
Figure~\ref{kacontprof}a shows overlapping contours, thus this division of the 
data, corresponding to a timescale of about 1 week, 
reveals no evidence for significant flux variability. 
Using the high and low-state data we then fit for the mean continuum slope 
as described in \S\ref{meansp} and overlaid the data in the K$\alpha$ band 
relative to the local power-law slope in each case. 
The two line profiles are shown as data/model ratios in 
figure~\ref{kacontprof}b and they appear indistinguishable. 
The same test, performed with the soft- and hard-state division 
discussed in \S\ref{humpfits}, leads us to conclude that 
no changes were observed either in flux or in shape of the line. 
Figure~\ref{kacontprof}c,d show the  confidence contour and the line profile, 
respectively, for the soft- and hard-state spectra. 
We further tested the possibility that the shape of the line might be variable,
on a $\sim$ 1 day timescale by overplotting the line profiles for spectra that 
looked qualitatively different. 
No significant changes in the line profile are observed. 

\vspace{-1.5cm}

\centerline{\includegraphics[width=9.5cm,height=11cm]{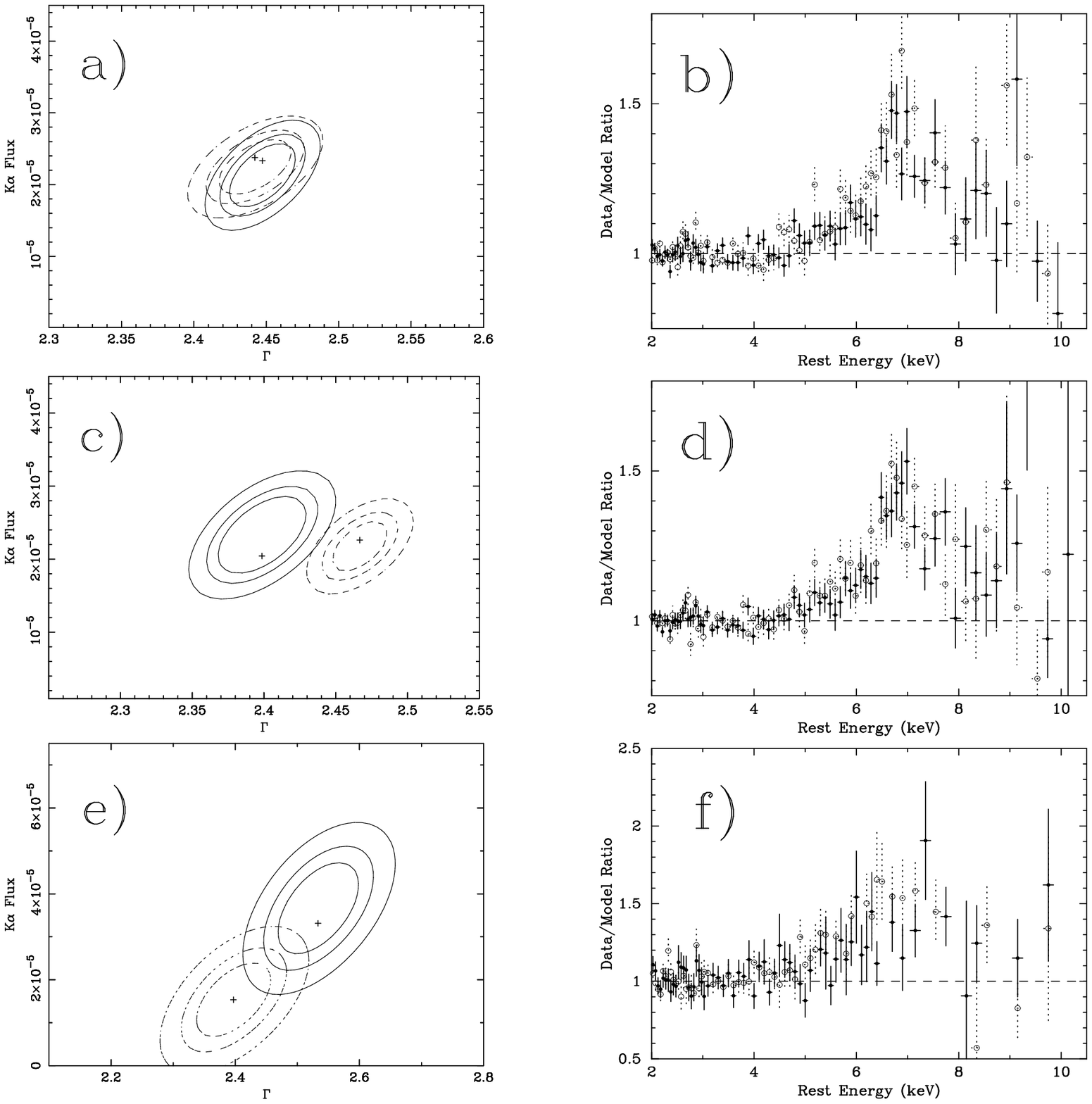}}

\vspace{-1.5cm}

\figcaption[Kalpha line]{Panel a:  
the $\Delta \chi^2 = 2.3, 4.61, 9.21$ contour levels for
Fe K-shell line intensity (in units of \phflux) vs.\ photon index $\Gamma$;
the full contours correspond to the high state (\S\ref{kafits}),
the dashed contours to the low state, 
and crosses indicate the best-fit values.
Panel b: the Fe K$\alpha$ regime compared to the continuum model; with overlay of 
high-state (open circles) and low states (filled circles). 
Panel c: same as a, with full contours corresponding to the hard state 
(\S\ref{humpfits}), dashed contours to the soft state. 
Panel d: same as b, with hard-state (open circles) and soft-state 
(filled circles) overlaid. 
Panel e: same as a, with full contours corresponding to the 
	high K$\alpha$ flux bin 4 (\S\ref{kafits}), the dashed 
	contours to the low K$\alpha$ flux bin 13. 
Panel f: same as b, with  bin 4 (open circles) and bin 13 (filled circles) overlaid.
\label{kacontprof}}
\centerline{}

\vspace{0.5cm}

Once assured that variations of the line shape, if present, are minor, 
we returned to our 14 time-selected spectra, where we considered the following 
models for K$\alpha$ line: 
(1) a {\tt laor} (Kerr) model; 
(2) a 2-Gaussian model (\S\ref{meanfek}) with the two fluxes 
	allowed to vary independently; 
(3) a 2-Gaussian model in which either the broad Gaussian flux  or 
(4) the narrow Gaussian is flux fixed to the best-fit 
	value from the mean spectrum; and 
(5) a 2-Gaussian model in which fluxes have fixed ratio derived from fits to 
	the mean spectrum. 
All fits were performed with photon index and the flux of the power law 
continuum left as free parameters. The shape of the K$\alpha$ line was 
kept fixed to the best fit values obtained for the mean spectrum. 
No model component was included for the soft hump, as we excluded the  
soft-band data for these fits. 
Figure~\ref{14fits} shows the time series for the K$\alpha$ line for 
the Kerr model, yielding results similar to those obtained from 
models (2) through (5). 
Fits to a constant model yield $\chi^2=3$ for 13 $dof$, 
thus the line does not show significant changes in flux when sampled on 
$\sim$ 1 day timescale. Analogous conclusions can be drawn for all other 
models considered. 

Finally, we considered the spectra extracted in bins 4 and 13,
which have the highest and the lowest measured K$\alpha$ flux, 
respectively (Figure~\ref{14fits}). 
We calculated confidence contours of line flux versus $\Gamma$ and overplotted 
the line profiles as we did above. Figure~\ref{kacontprof}e
shows the contour plots which demonstrate that the flux variation is 
insignificant and there is a large overlap of the contours. 
Figure~\ref{kacontprof}f shows the line profiles, whose large equivalent widths
are consistent at 90\,\% confidence (EW$_{\rm 4} = 889^{+219}_{-368}$\,eV, 
EW$_{\rm 13} = 411^{+252}_{-249}$\,eV).
Therefore, we do not detect significant variations in EW or flux of the 
K$\alpha$ line. With the  signal-to-noise available using these data 
the line flux would have had to vary by at least a factor of 2 
for us to detect it at the 99\,\% confidence level 
(cf.\ Figure~\ref{kacontprof}e). 

We also applied a {\tt laor} model to the 1996 \ASCA{} observation 
\citep{TGN98}, within the same energy ranges as we used here and with the  
corrections for the SIS low-energy degradation appropriate to 1996
(see \S\ref{SIStrouble}), 
and obtained peak energy, EW, and inclination for the K$\alpha$ 
line consistent with what we found in \S\ref{meanfek}.

	\subsection{RMS Spectra\label{rms}}  

In optical/UV spectroscopy of active galactic nuclei, 
it is easy to obtain a series of spectra of sufficiently high 
S/N to perform 
time-resolved spectroscopic analysis. In this case, 
a useful way to isolate variable features is to construct a 
root-mean-square (rms) spectrum. 
This ASCA long look has allowed us to use our 14 
individual time-selected spectra described in \S\ref{ciccio} 
in analogous way.  For each detector 
the 14 spectra have been rebinned so that each bin has at least a 5-sigma 
detection (but up to a maximum of 10 adjacent bins were combined 
to achieve a signal-to-noise ratio of 5 within a bin) 
then they were degraded to the resolution of the worst spectrum. 
We created rms and mean spectra by calculating the  
rms and mean flux in each bin. The choice of a simple mean as 
opposed to a weighted mean was dictated by the fact that we did 
not want to favor the high state spectra. 
Figure~\ref{rms_on_mean} shows the ratio rms/mean spectrum for SIS1 and SIS0, 
using data with energy $>0.7$\,keV. The mean value is $\sim 20$\,\% 
and there is no systematic trend for different energies. 
This is consistent with the results from \S\ref{meanhump} 
showing that the power-law continuum component contributes most of 
the flux in the hard band and $\sim 72$\,\% of the flux in 
the soft band.
The possible weak trend for energies $< 2$\,keV of increased variability 
for decreasing energy, may suggest that the soft hump is partially driving 
the variability in the soft band. 
The ``spike-like'' features at $\sim$\,1.7--2.5\,keV are probably 
due to calibration uncertainties.

\centerline{\includegraphics[width=8.5cm]{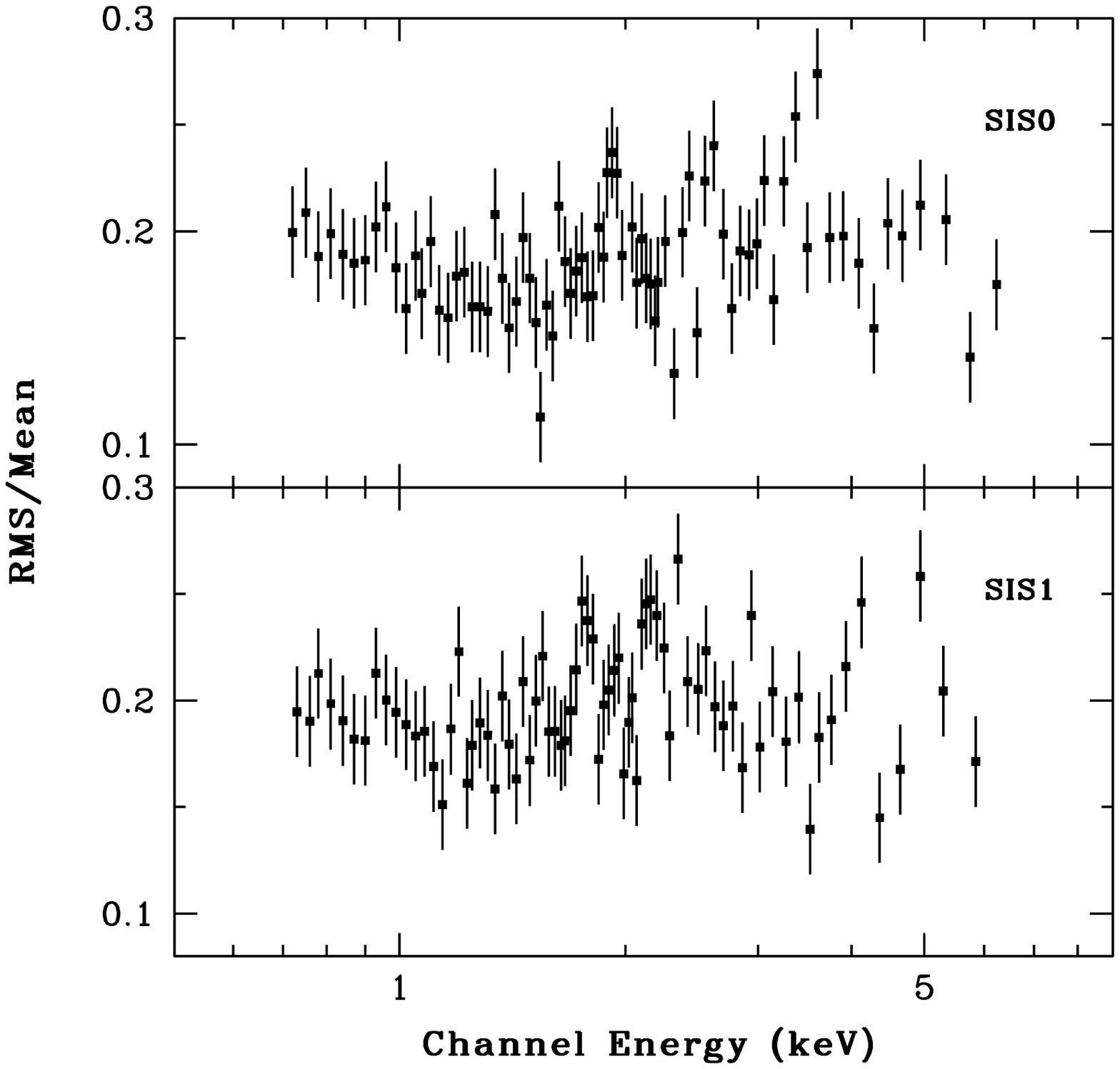}}
\figcaption[RMS/Mean spectrum]{The ratio of RMS/Mean
spectrum for SIS1 and SIS0, using data of energy $>0.7$\,keV.
The ``spike-like'' features at $\sim$\,1.7--2.5\,keV are probably 
due to calibration uncertainties. 
\label{rms_on_mean}}
\centerline{}

\section{Summary of Observational Results\label{results}} 

\begin{enumerate}
\item   On a 12-day baseline, the X-ray flux of Ton~S180 presents 
	trough-to-peak variations by a factor 3.5 in the 0.7--1.3\,keV band, 
	and 3.9 in the 2--10\,keV band when sampled using 256\,s bins. 
\item   The mean photon index, calculated from the continuum fit of the 
	mean spectrum, is $\Gamma= 2.44\pm0.02$. Time-resolved spectroscopy
	reveals significant changes on timescales of $\sim$ 1 week. 
\item   We confirm the presence of a separate ``soft hump'' component 
	at energies $<2$\,keV. This component shows flux variations 
	down to timescales of $\sim$ 1 day, ranging by a factor of 2.3 
	in normalization over the 12 days of our observations. 
	Some flux changes appear to match changes in the 2--10\,keV flux on this 
	timescale although the soft hump shows a drop in EW from 100 to 84\,eV 
	across the observation. 
	Minor changes in the shape of the soft hump are apparent on timescales 
	of a few days. 
\item   In our approximately daily sampling the photon index is correlated 
	with the flux of the soft X-ray hump. 
\item   Variations in both the photon index and soft hump strength 
	contribute to the change in softness ratio observed across the 
	\ASCA{} observation. 
\item   The softness ratio reveals spectral variability down to timescales 
	of $\sim 1000$\,s in addition to the slow decline across the observation.
	The power-law and soft hump show divergent behavior on very short 
	(1000\,s) or long (2 weeks) timescales. 	 
\item   The Fe K$\alpha$ line emission is detected with a narrow component
	peaking at $\sim$ 6.8\,keV, indicating an origin in ionized material. 
	A broad line component (EW $\sim 500$\,eV) is also evident. 
	We do not detect significant variations of the Fe K$\alpha$ line 
	strength or equivalent width, on timescales of $\sim$ 1 day--12 days.  
\end{enumerate}

\section{Comparison with Akn~564\label{comptonakn}}

Comparison of our results for Ton~S180 with those obtained for Akn~564 
\citep{Akn564I} reveals broad similarity in the overall shape of the 
X-ray spectra (steep power-law continuum, strong soft excess, and ionized, 
large EW Fe K$\alpha$), and variability characteristics.  
Given these similarities, that may be indicative of the 
characteristics of the NLS1 as a class, 
we summarize the properties of these two NLS1s in Table~\ref{tonakn}. 
Column (1) lists the properties; Column (2) and (3) the values for 
Ton~S180 and Akn~564, respectively; Column (4) and (5) the references for 
Ton~S180 and Akn~564, respectively.

   \section{Discussion and Conclusions \label{discussion}} 

Ton~S180 reaffirms the rapid and large-amplitude X-ray variations which are a 
characteristic of the NLS1 class. During our 12-day observation, 
the light curves sampled on 256\,s timescales show trough-to-peak flux 
variations by a factor of $\sim$ 3.5 in the soft band (0.7-1.3\,keV), and 
up to a factor of $\sim$ 4 in the hard band (2--10\,keV). 
Further examination of these light curves shows variations up to a factor of
$\sim 2$ occurring on timescales of $\sim$ 1000 s which are sometimes 
accompanied by spectral changes. There is a strong  correlation
($r_{\rm max} = 0.748$) between the soft- and hard-band light curves, with a 
95\,\% confidence  upper limit on the lag of hard with respect to soft of 
$\tau_{\rm cent} < 0.07$\,d. 
The strength of this correlation, the short lag, and the fact that 
$\sim$ 72\,\% of the flux in the soft band is contributed by the 
power-law continuum  while the hard flux is driven by the power-law, 
indicate that this correlation is dominated by the variations of the 
power-law continuum itself. 
	
We used these timing data to obtain an estimate of some fundamental 
parameters for Ton~S180. 
Our estimate of the accretion efficiency is $\eta \gtsim 7$\,\%, 
marginally above the limit for the efficiency in the Schwarzschild geometry,  
perhaps favoring a Kerr geometry. However, large uncertainties are 
associated with this estimate \citep[see][]{Brandtea99}. 
Light-crossing time arguments yield a radius $R$ for the emitting region, 
$R \lesssim 12 \; R_{\rm S}$ for $M_{\rm BH} \gtsim 8 \times 10^{6} M_{\odot}$.
The 1000\,s timescale for variability observed in Ton~S180 indicates that  
neither thermal nor viscous/radial drift phenomena can be held responsible 
for the observed fast variability. The values of these parameters are 
consistent with what is generally found for other Seyfert 1s.
\citet{Haardtea94} discuss a model where soft thermal photons from the 
accretion disk are Compton-upscattered in localized blobs 
of coronal plasma, constrained by magnetic loops from the disk. 
The resulting X-rays from the blobs then illuminate the disk and 
produce the so-called Compton Hump and Fe K$\alpha$ emission line. 
The amplitudes and timescales of the rapid variations observed 
in Ton~S180 are consistent with those expected as a result of 
stochastic noise in the number of reprocessing blobs, which 
depend on the formation and reconnection of magnetic loops. 

The continuum fit to the mean spectrum yields a photon index 
$\Gamma = 2.44 \pm 0.02$. Our time-selected spectral fits 
give $\langle \Gamma \rangle = 2.47 \pm 0.07$ (1-$\sigma$ error) 
with variations from 2.38 to 2.62. 
The steep index and the range ($\Delta\Gamma = 0.24$) are comparable 
with that observed in Akn~564 \citep[$\Gamma=2.54$, 
$\Delta\Gamma = 0.27$,][]{Akn564I},
though the variations on a $\sim 1$\,day timescale are significant
in the case of Akn~564 and not for Ton~S180 
(where we detect variations of $\Gamma$ in $\sim 1$\,week timescale). 
The range of indices is also consistent with that found for the BLS1 
NGC~7469 \citep[$\Delta\Gamma = 0.32$,][]{Nandraea2000}. 
This may indicate a fundamental similarity in the process which causes 
slope fluctuations, over a range of different AGN. 

We confirm the presence of the soft X-ray emission component at energies 
$<2$\,keV previously observed by \citet{Finkea97}, \citet{Comastriea98}, 
and \citet{TGN98}. 
This component rises above the power-law continuum 
$\lesssim 2$\,keV, and has recently been shown to be a smooth 
continuum component rather than a blend of features, not only in Ton~S180 
\citep{Turnerea01}, but also in the NLS1 NGC~4051 \citep{Collea01}. 
Therefore, our time-selected spectroscopic analysis is particularly well 
suited to study its variability properties in particular and 
understand the nature of the soft hump in NLS1s in general. 
The soft hump shows flux variations on timescales as short as $\sim 1$\,day 
(the shortest timescales our time-selected spectroscopy can study), 
ranging by a factor of 2.3 in flux over the observation, but always being present. 
We note that the fast variability observed in the soft hump of Ton~S180 
rules out an origin of the soft emission 
in large-scale components, such as circumnuclear starburst (as 
also concluded for the soft hump in Akn~564, \citealt{Akn564I}).
The hard flux, when similarly binned on timescales of a day, has a range of a 
factor 1.65. 	
The softness ratio shows spectral variability on timescales as short as 
$\sim 1000$\,s. In Ton~S180 such fast spectral variability can be 
attributed either to rapid changes in photon index or in the relative 
strengths of the soft hump and power-law. 
In addition to this evidence for divergent behavior on very short timescales, 
the soft hump and power-law show different trends on timescales of $\sim 1$ week.  
The higher value of $F_{\rm var}$ measured in the soft than in the hard 
band is probably due to the combined effects of a larger 
amplitude of variation of the soft hump over the baselines of the 
observation, and the changes in spectral slope (which could preferentially 
affect the soft band if the pivot point lies within the hard band). 
The correlation observed between the soft hump flux and the 
photon index in Ton~S180 is expected in disk-corona models 
where an increase of the flux in the soft X-ray/UV component 
can cool the corona and steepen the power-law continuum
\citep{Haardtea94,PDO95}. 
This is not the case for Akn~564, where \citet{Akn564I} 
suggest the corona cooling might be saturated.

Iron K$\alpha$ line emission is detected, with a broad, asymmetric profile.  
It is best parameterized by two Gaussian components; the narrow component 
peaks at $E_{\rm N}=6.81$\,keV, which is consistent with emission from 
highly ionized gas, the broad component consistent with emission from either 
neutral or ionized gas. We caution about the reliability of this result
because the signal-to-noise in this range is low and the errors on the fitted
parameters are correspondingly large (Table~\ref{kafitstab}). 
Therefore it seems that the best parametrization for the Fe K$\alpha$ line
is a broad ionized line, which is consistent with previous 
\BeppoSAX{} \citep{Comastriea98} and \ASCA{} \citep{TGN98} observations.
The broad component is generally thought to originate in the innermost 
regions of the accretion disk around the central black hole \citep[e.g.,][and 
references therein]{Fabianea00}, and so it is due to highly ionized Fe, 
while the narrow component is thought to be generated in the putative obscuring 
torus that extends on a parsec scale \citep[][]{GHM94,KMZ94,Yaqoobea01}, 
and so is believed to be due to neutral material. 
In this context, the result of an ionized narrow Fe K$\alpha$ component is therefore 
somewhat suprising. Note, however, that observations of such a feature are 
 not unprecedented: \citet{Sakoea00} and \citet{Ogleea00} detected 
narrow lines due to highly ionized iron in \Chandra{} observations of 
Mkn~3 and NGC~4151, respectively. 

The Fe K$\alpha$ emission line has a very large equivalent width, 
EW $\approx 500$\,eV, as previously observed in other \ASCA{}  
and {\it BeppoSAX} observations of Ton~S180, Akn~564,
and other NLS1s \citep{TGN98,Comastriea98,Akn564I}. 
These EWs are interesting compared to a sample taken across the Seyfert 
1 population \citep{Nandraea97b}, which found an average EW $= 230$\,eV.
One interpretation is in terms of an extreme Fe abundance in NLS1s, 
as proposed by \citet{TGN99} for Akn~564; this may, in turn, 
support the proposition that NLS1s are Seyfert galaxies in an early 
stage of evolution \citep{Mathur00}. 
Alternatively, \citet{MFR96} show that the expected EWs for predominantly 
H- and He-like Fe can be a factor of a few higher than those from neutral Fe. 
We do not detect significant variations of the Fe K$\alpha$ line, 
on timescales of $\sim$ 1 day or of $\sim$ 1 week.
We note that given our signal-to-noise in the K$\alpha$ 
energy range for the 14 time-selected spectra, the line flux would 
have had to varied by at least  a factor 2 for us to have detected the 
variation at $> 99$\,\% confidence level.

\acknowledgements 

PR acknowledges support through NASA ADP grant NAG5-9346-1.  
TJT is pleased to acknowledge support for this work by NASA through  
grant number NAG5-7385 (LTSA). 
SM acknowledges support through NASA grant NAG5-8913 (LTSA).
We also acknowledge support from  HST--GO--08265.01--A from the 
Space Science Telescope Institute, which is operated by the Association of 
Universities for Research in Astronomy, Inc., under NASA contract 
NSS5-226555. 
We thank the \ASCA{} team for their operation of the  
satellite, and Tahir Yaqoob for discussions on the \ASCA{}  
calibration. We thank Brad Peterson for use of the cross-correlation 
codes and for comments and Rick Pogge for careful reading of the 
manuscript. We are grateful to the anonymous referee for making some
important suggestions.
This research has made use of the NASA/IPAC Extragalactic Database 
(NED) which is operated by the Jet Propulsion Laboratory, California 
Institute of Technology, under contract with the National Aeronautics 
and Space Administration.

 \vspace{-0.9truecm}
\begin{deluxetable}{lllcc}        
 \tablewidth{510pt}
 \vspace{-0.9truecm}
 \tablecaption{Comparison of the Properties of Ton~S180 and Akn~564.\label{tonakn}}
\tablehead{ \colhead{} & \colhead{Ton~S180\tablenotemark{a}} & \colhead{Akn~564\tablenotemark{a}} 
			& \multicolumn{2}{c}{References\tablenotemark{b}}\\
            \colhead{(1)} & \colhead{(2)} & \colhead{(3)} & \colhead{(4)} & \colhead{(5)} }
\startdata
$z$                     		& 0.06198      		& 0.0247  		& 1 & 2 \\
Galactic $N_{\rm H}$ (cm$^{-2}$) 	& 1.52$\times 10^{20}$ & 6.4$\times 10^{20}$ 	& 3 & 4 \\
$V$ (mag)                 		& 14.4         		& 14.6 	 		& 5 & 6 \\
$L_{2-10}$ (10$^{43}$ \ergsec)        	& 4.9       		& 2.4  			& 7(3) & 8(3) \\
$R_{\rm max}$ {} soft band  		& 3.5       		& 16 			& 7(3) & 8(3) 			\\*
$R_{\rm max}$  hard band 		& 3.9          		& 14 			& 7(3) & 8(3) \\
$R_{\rm max}$  Soft Hump		& 2.33 $\pm$ 0.45      	& 6.44 $\pm$ 3.30   	& 7(5.3) & 8(5.3) \\
$R_{\rm max}$  2--10\,keV 		& 1.65 $\pm$ 0.02      	& 3.97 $\pm$ 0.06   	& 7(5.3) & 8(5.3) \\
$F_{\rm var}$ (12 d, soft) (\%) 	& 19.12 $\pm$ 0.58   	& 35.81 $\pm$ 0.81 	& 7(3.1) & 7 \\
$F_{\rm var}$ (12 d, hard) (\%) 	& 17.26 $\pm$ 0.65   	& 35.73 $\pm$ 0.70     	& 7(3.1) & 7 \\
$F_{\rm var}$ (1 d, soft) (\%)         	& variable              & variable  	 	& 7(3.1) & 8(3.1) \\
\hspace{0.3cm}correlates with X-ray     & no                   	& no  			& 7(3.1) & 8(3.1,7) \\
 $F_{\rm var}$ (1 d, hard) (\%)      	& variable, $<$ soft  	& variable, $<$ soft  	& 7(3.1) & 8(3.1) \\
\hspace{0.3cm}correlates with X-ray     & no                   	& no  			& 7(3.1) & 8(3.1,7) \\
ICCF $r_{\rm max}$ (soft--hard band)	& 0.748        		& 0.942   		& 7(3) & 8(7) \\
ICCF $\tau_{\rm cent}$ (95\% limit, 
d)\tablenotemark{c} 			& $< 0.07$   		& $<$ 0.02  		& 7(3) & 8(7) \\
$\Gamma$ (Mean Spectrum)  		& 2.44 $\pm$0.02  	& 2.538 $\pm$ 0.005  	& 7(4) & 8(4) \\
\hspace{0.3cm}
$\Gamma_{\rm min}$--$\Gamma_{\rm max}$  & 2.38--2.62   		& 2.45--2.72 		& 7(5.2) & 8(5.2) \\
\hspace{0.3cm}$\Delta \Gamma$      	& 0.24         		& 0.27  		& 7(5.2) & 8(5.2) \\
PL soft band contribution (\%)  	& 72            	& 75    		& 7(4.1) & 8(3.1) \\
Soft Hump Gaussian Fits: $E$ (keV)         		& 0.17 $\pm$ 0.17    	 & 0.57 $\pm$ 0.2  & 7(4.1) & 8(4.1) \\
\phm{Soft Hump Gaussian  } FWHM (keV)     	& 1.01$^{+0.05}_{-0.12}$ & 0.36 $\pm$ 0.01  	   & 7 & 8 \\
\phm{Soft Hump Gaussian  } $n$ ($10^{-2}$ 
	ph s$^{-1}$ cm$^{-2}$)  			& 1.50$^{+0.37}_{-0.61}$ & 1.25$^{+0.12}_{-0.17}$ & 7 & 8 \\
\phm{Soft Hump Gaussian  } EW (eV)          	& 94$^{+23}_{-38}$       & 110$^{+11}_{-15}$  	& 7 & 8 \\
K$\alpha$ {\tt diskline} Fits: $E$ (keV)\tablenotemark{d} & $6.40^{+0.27}_{-0.00p}$ & $7.00^{+0.00}_{-0.13}$ & 7(4.2) & 8(4.2) \\
\phm{K$\alpha$ {\tt diskline} Fits:} EW (eV)        	& $461^{+120}_{-84}$     & 351$\pm$ 85  		& 7 & 8 \\
\phm{K$\alpha$ {\tt diskline} Fits:} $i$ (deg)        	& $35^{+22}_{-35p}$     & 26$\pm$2  		& 7 & 8 \\
K$\alpha$ {\tt laor} Fits:{}{} 
		$E$ (keV)\tablenotemark{e}   		& $6.55^{+0.16}_{-0.15p}$ & $6.99^{+0.01p}_{-0.13}$ & 7(4.2) & 8(4.2) \\
\phm{K$\alpha$ {\tt laor} Fits:{}{}{}} EW (eV)        	& $517^{+123}_{-111}$ 	& $653\pm85$   		& 7 & 8 \\
\phm{K$\alpha$ {\tt laor} Fits:{}{}{}} $i$ (deg)        & $23^{+14}_{-23p}$    	& $17^{+11}_{-17}$	& 7 & 8 \\
\enddata
\tablenotetext{a}{90\,\% confidence level uncertainties.} 
\tablenotetext{b}{The numbers in parenthesis refer to the sections where the relevant information
is found in the given references.}
\tablenotetext{c}{Lag of hard relative to soft flux.}
\tablenotetext{d}{Best parametrization for Ton~S180 is by two Gaussians with peak energy 
$E_{\rm N}=6.81^{+0.08}_{-0.12}$\,keV and $E_{\rm B}=6.58^{+0.25}_{-0.28}$\,keV.}
\tablenotetext{e}{Best-fit model for Akn~564 \citep{Akn564I}.}
\tablerefs{1: \citet{Wisotzkiea95}; 2: \citet{Huchraea99}; 3: 
\citet{Starkea92}; 4: \citet{DickeyL90}; 5:  NED; 6: \citet{rc3.9catalogue}; 7:  
This work; 8: \citet{Akn564I}. }

\end{deluxetable}       


\begin{thebibliography}{} 	
\bibitem[Ballantyne, Iwasawa, \& Fabian(2001)]{BIF01}
	Ballantyne, D.\ R., Iwasawa, K., \& Fabian, A.\ C.\  2001,
	\mnras, 323, 506 
\bibitem[Boller, Brandt, \& Fink(1996)]{BBF96} 
	Boller, Th., Brandt, W.\ N., \& Fink, H.\ 1996, \aap, 305, 53 
\bibitem[Boroson \& Green(1992)]{BG92} 
	Boroson, T.\ A. \& Green, R.\ F. 1992, \apjs, 80, 109 
\bibitem[Brandt, Mathur, \& Elvis(1997)]{BME97}
	Brandt, W.\ N., Mathur, S.,  \& Elvis, M. 1997, \mnras, 
	285, L25
\bibitem[Brandt et al.(1999)]{Brandtea99} 
	Brandt, W.\ N., Boller, T., Fabian, A.\ C., \& Ruszkowski, M.\ 1999, 
	\mnras, 303, L53 
\bibitem[Burke et al.(1991)]{Burkeea91} 
	Burke, B.\ E., Mountain, R.\ W., Harrison, D.\ C., Bautz,  
	M.\ W., Doty, J.\ P., Ricker, G.\ R., \& Daniels, P.\ J. 
	1991, IEEE Trans. ED-38, 1069 
\bibitem[Collinge et al.(2001)]{Collea01} 
	Collinge, M.\ J., Brandt, W.\ N., Kaspi, S., Crenshaw, D.\ M.,  
	Elvis, M., Kraemer, S.\ B., Reynolds, C.\ S., Sambruna, R.,  \& 
	Wills, B.\ 2001, \apj, in press (astro-ph/0104125)
\bibitem[Comastri et al.(1998)]{Comastriea98}		 
	Comastri, A., et al.\  1998, \aap, 333, 31 
\bibitem[Comastri et al.(2001)]{Comastriea01}		
	Comastri, A., et al.\ 2001, \aap, 365, 400
\bibitem[de Vaucouleurs et al.(1991)]{rc3.9catalogue}
        de Vaucouleurs, G., de Vaucouleurs, A., Corwin, H.\ G.,
        Buta, R.\ J., Paturel, G., \& Fouque, P.\ 1991, \skytel, 82, 621 
\bibitem[Dickey \& Lockman(1990)]{DickeyL90}
        Dickey, J.\ M., \& Lockman, F.\ M. 1990, \araa, 28, 215  
\bibitem[Edelson et al.(2001)]{Edelsonea01} 
	Edelson, R.,  et al.\ 2001, submitted 
\bibitem[Fabian et al.(1989)]{Fabianea89} 
	Fabian, A.\ C., Rees, M.\ J., Stella, L., \&  
	White, N.\ E. 1989, \mnras, 238, 729 
\bibitem[Fabian et al.(2000)]{Fabianea00} 
	Fabian, A.\ C., Iwasawa, K., Reynolds, C.\ S., \& Young, A.\ J.\ 
	2000, \pasp, 112, 1145 
\bibitem[Fink et al.(1997)]{Finkea97} 
	Fink, H.\ H., Walter, R., Schartel, N., \& 
	Engels, D.\ 1997, \aap, 317, 25 
\bibitem[Gaskell \& Sparke(1986)]{GasSpar86} 
	Gaskell, C.\ M., \& Sparke, L.\ S.\ 1986, \apj, 305, 175 
\bibitem[Gaskell \& Peterson(1987)]{GasPet87} 
	Gaskell, C.\ M., \& Peterson, B.\ M.\ 1987, \apjs, 65, 1 
\bibitem[Ghisellini, Haardt, \& Matt(1994)]{GHM94} 
	Ghisellini, G., Haardt, F., \& Matt, G.\ 1994, \mnras, 267, 743 
\bibitem[Goodrich(1989)]{Goodrich89} 
	Goodrich, R.\ W.\  1989, \apj, 342, 234  
\bibitem[Guainazzi et al.(1998)]{Guainazziea98}
	Guainazzi, M., Piro, L., Capalbi, M., Parmar, A.\ N.,
	Yamaguchi, M., \& Matuoka, M. 1998, \aap, 339, 327
\bibitem[Guilbert, Fabian, \& McCray(1983)]{GFM83} 
	Guilbert, P.\ W., Fabian, A.\ C., \& McCray, R. 1983, \apj, 266, 466 
\bibitem[Haardt, Maraschi, \& Ghisellini(1994)]{Haardtea94} 
	Haardt, F., Maraschi, L., \& Ghisellini, G.\ 1994, \apj, 432, L95 
\bibitem[Huchra, Vogeley \& Geller(1999)]{Huchraea99}
        Huchra, J.\ P., Vogeley, M.\ S., \& Geller, M.\ J.\
        1999, \apjs, 121, 287  
\bibitem[Kaspi et al.(2000)]{K2000} 
	Kaspi, S., Smith, P.\ S., Netzer, H., Maoz, D., Jannuzi, B.\ T., \&  
	Giveon, U.\ 2000, \apj, 533, 631  
\bibitem[Krolik, Madau, \& {\. Z}ycki(1994)]{KMZ94} 
	Krolik, J.\ H., Madau, P., \& {\. Z}ycki, P.\ T.\ 1994, \apj, 420, L57 
\bibitem[Kuraszkiewicz et al.(2000)]{Joannaea00} 
	Kuraszkiewicz, J., Wilkes, B.\ J., Brandt, W.\ N., \& Vestergaard, M.\ 
	2000, \apj, 542, 631
\bibitem[Laor et al.(1997)]{Laorea97} 
	Laor, A., Fiore, F., Elvis, M., Wilkes, B.\ J., 
	\& McDowell, J.\ C.\ 1997, \apj, 477, 93 
\bibitem[Laor(1991)]{Laor91} 
	Laor, A.\ 1991, \apj, 376, 90
\bibitem[Leighly(1999a)]{Leighly99I}		
	Leighly, K.\ M.\ 1999a, \apjs, 125, 297 
\bibitem[Leighly(1999b)]{Leighly99II}		
	Leighly, K.\ M.\ 1999b, \apjs, 125, 317  
\bibitem[Mason, Puchnarewicz, \& Jones(1996)]{MPJ96}
	Mason, K.\ O., Puchnarewicz, E.\ M., \& Jones, L.\ R.\ 1996, \mnras, 283, L26
\bibitem[Mathur(2000)]{Mathur00} 
	Mathur, S.\ 2000, \mnras, 314, L17 
\bibitem[Matt, Fabian, \& Ross(1993)]{MFR93}
	Matt, G., Fabian, A.\ C.,  \& Ross, R.\ R.\ 1993, 
	\mnras, 264, 839
\bibitem[Matt, Fabian, \& Ross(1996)]{MFR96}
	Matt, G., Fabian, A.\ C.,  \& Ross, R.\ R.\ 1996, 
	\mnras, 278, 1111
\bibitem[Nandra et al.(1997a)]{Nandraea97a} 
 	Nandra, K., George, I.\ M., Mushotzky, R.\ F., Turner, T.\ J., 
	\& Yaqoob, T.\ 1997a, \apj, 476, 70 
\bibitem[Nandra et al.(1997b)]{Nandraea97b} 
 	Nandra, K., George, I.\ M., Mushotzky, R.\ F., Turner, T.\ J., 
	\& Yaqoob, T.\ 1997b, \apj, 477, 602 
\bibitem[Nandra et al.(2000)]{Nandraea2000}  
	Nandra, K., Le, T., George, I.\ M., Edelson, R.\ A.,  
	Mushotzky, R.\ F., Peterson, B.\ M., \& Turner, T.\ J.\  
	2000, \apj, 544, 734  
\bibitem[Ogle et al.(2000)]{Ogleea00}
	Ogle, P.\ M., Marshall, H.\ L., Lee, J.\ C., \& Canizares, C.\ R.\ 
	2000, \apjl, 545, L81 
\bibitem[Ohashi et al.(1996)]{Ohashiea96} 
	Ohashi, T., et al.\ 1996, \pasj, 48, 157 
\bibitem[Osterbrock \& Pogge(1985)]{OP85} 
 	Osterbrock, D.\ E., \&  Pogge, R.\ W. 1985, \apj, 297, 166  
\bibitem[Peterson(1993)]{P93} 
	Peterson, B.\ M. 1993, PASP, 105, 207 
\bibitem[Peterson et al.(1998)]{Petersonea98} 
	Peterson, B. M., Wanders, I., Horne, K., Collier, S.,  
	Alexander, T., Kaspi, S., \& Maoz, D.\ 1998, \pasp, 110, 660  
\bibitem[Peterson et al.(2000)]{P2000} 
	Peterson, B. M. et al. 2000, \apj, 542, 161 
\bibitem[Pounds, Done, \& Osborne(1995)]{PDO95}
	Pounds, K.\ A., Done, C., \& Osborne, J.\ P.\ 1995, \mnras, 
	277, L5
\bibitem[Sako et al.(2000)]{Sakoea00} 
	Sako, M., Kahn, S.\ M., Paerels, F., \& Liedahl, D.\ A.\ 
	2000, \apj, 543, L115 
\bibitem[Stark et al.(1992)]{Starkea92}
	Stark, A.\ A., Gammie, C.\ F., Wilson, R.\ W., Bally, J., 
	Linke, R.\ A., Heiles, C., \& Hurwitz, M.\  1992, ApJS, 79, 77
\bibitem[Turner, George, \& Nandra(1998)]{TGN98} 	
  	Turner, T.\ J., George, I.\ M., \& Nandra, K.\ 1998, \apj,  
	508, 648
\bibitem[Turner, George, \& Netzer(1999)]{TGN99} 	
 	Turner, T.\ J., George, I.\ M., \& Netzer, H.\ 1999, \apj,  
	526, 52 
\bibitem[Turner et al.(1999)]{Turnerea99b} 
	Turner, T.\ J., et al.\ 1999, in Proceedings of the 19th  
	Texas Symposium on Relativistic  Astrophysics and Cosmology,  
	ed. J. Paul, T. Montmerle, \& E. Aubourg (Saclay: CEA), E441 
\bibitem[Turner et al.(2001a)]{Turnerea01}  
	Turner, T.\ J., et al.\  2001a, \apj, 548, L13 
\bibitem[Turner et al.(2001b)]{Akn564I}
	Turner, T.\ J., Romano, P., George, I.\ M., Edelson, R., 
	Collier, S.\ J., Mathur, S., \& Peterson, B.\ M.\ 2001b,
	\apj, in press (astro-ph/0105238)
\bibitem[Turner et al.(2001c)]{Turnerea01c}  		
	Turner, T.\ J., et al.\  2001c, in preparation 
\bibitem[Vaughan et al.(1999a)]{Vaughan99a} 	 
 	Vaughan, S., Pounds, K.\ A., Reeves, J., Warwick, R.,
 	\& Edelson, R.\ 1999, \mnras, 308, L34
\bibitem[Vaughan et al.(1999b)]{Vaughan99b} 		
	Vaughan, S., Reeves, J., Warwick, R., \& Edelson, R.\ 
	1999, \mnras, 309, 113
\bibitem[Wandel \& Boller(1998)]{WB98}
	Wandel, A., \& Boller, Th.\ 1998, \aap, 331, 884
\bibitem[White \& Peterson(1994)]{WP94} 
	White, R.\ J., \& Peterson, B.\ M.\ 1994, \pasp, 106, 879 
\bibitem[Wisotzki et al.(1995)]{Wisotzkiea95}
	Wisotzki, A., Dreizler, S., Engels, D., Fink, H.\ H., \& 
	Heber, U.\ 1995, \aap, 297, L55
\bibitem[Yaqoob et al.(2000)]{Yaqoob}  
	Yaqoob, T., et al.\ 2000, \ASCA{} GOF Calibration Memo,  
	ASCA-CAL-00-06-01, v1.0 
\bibitem[Yaqoob et al.(2001)]{Yaqoobea01} 
	Yaqoob, T., George, I.\ M., Nandra, K., Turner, T.\ J., 
	Serlemitsos, P.\ J., \& Mushotzky, R.\ F.\ 2001, \apj, 546, 759 
\end{thebibliography}
\end{document}